\documentclass[aps,epsfig,prd,longbibliography]{revtex4-2}
\usepackage[english]{babel}

\usepackage[letterpaper,top=2cm,bottom=2cm,left=3cm,right=3cm,marginparwidth=1.75cm]{geometry}

\usepackage[usenames,dvipsnames]{color}
\usepackage{amsmath,mathtools}
\usepackage{graphicx}
\usepackage[colorlinks=true, allcolors=blue]{hyperref}
\usepackage{latexsym}
\usepackage{amsfonts}
\usepackage{amssymb}
\usepackage{array}
\usepackage{color}
\usepackage{subfigure}
\usepackage{verbatim}
\usepackage{ulem}
\usepackage{comment}
\usepackage{bm}
\usepackage{braket} 
\usepackage{xcolor}
\usepackage[utf8]{inputenc} % Specifies the input encoding is UTF-8
\usepackage[T1]{fontenc}      % Selects the T1 font encoding for proper output
\usepackage{placeins} 
\usepackage{appendix}
\newcommand{\braketk}[2]{\left\langle#1|#2\right\rangle}

\newcommand{\av}[1]{\langle#1\rangle}
\newcommand{\ii}{\mathrm{i}}

\newcommand{\cmmnt}[1]{}

\begin{document}
\title{Qubit-parity interference despite unknown interaction phases}
%\author{Kratveer Singh{These authors contributed equally.} \and Kimin Park\footnotemark[1] \and  Vojtech Svarc \and  Lukas Slodicka \and Radim Filip}
%\affiliation{Department of Optics, Palacky University, 77146 Olomouc, Czech Republic}

\author{Kratveer Singh}
\thanks{These authors contributed equally.}
\email{kratveer.singh@upol.cz;kratveer050@gmail.com}
\author{Kimin Park}
\thanks{These authors contributed equally.}
\email{park@optics.upol.cz}
\author{Vojt\v{e}ch \v{S}varc}
\author{Artem Kovalenko}
\author{Tuan Pham}
\author{Ond\v{r}ej \v{C}íp }
\author{Lukáš Slodička}
\email{slodicka@optics.upol.cz}
\author{Radim Filip}
\email{filip@optics.upol.cz}
\affiliation{Department of Optics, Palacky University, 77146 Olomouc, Czech Republic}

% \author[*]{Kratveer Singh}
% \author[*]{Kimin Park}
% \author[ ]{Vojtech Svarc}
% \author[ ]{Lukas Slodicka}
% \author[ ]{Radim Filip}
% \affil[*]{These authors contributed equally.}
% \affil[ ]{Department of Optics, Palacky University, 77146 Olomouc, Czech Republic}

\date{\today}

\begin{abstract}

Quantum interference between interacting systems is fundamental to basic science and quantum technology, but it typically requires precise control of the interaction phases of lasers or microwave generators. %but the interaction phase must typically  be {precisely} controlled {to observe} the interference. 
%It arises as a phase-sensitive effect between two interacting quantum subsystems. 
Can interference be observed if those interaction phases are stable but unknown, 
usually prohibitive for complex state without active control?
Here, we answer this question  by experimentally preparing a  Schr\"odinger-cat-like state of an internal qubit and a motional oscillator of a trapped $^{40}$Ca$^{+}$ ion, and its {robustness}  to such uncontrolled {phases}.
%Our test, using a sequence of alternating red and blue sideband pulses, enforces a strict qubit-parity correlation  inherently insensitive 
By applying alternating red and blue sideband pulses, we enforce a strict qubit-parity correlation {and interference} inherently insensitive to stable but unknown phases of the driving laser. 
%Despite this {stable but} unknown phase {which would completely {erase} a conventional protocol's interference fringes, we show the underlying coherence structure remains intact. }
%We first demonstrate the presence of coherence using a single-pulse probe, which distinguishes the prepared state from an incoherent mixture. 
{For this qubit-parity interference, } we use a {minimal} two-pulse interferometric sequence to %{implement an orthogonal phase mapping for direct sensing of hybrid qubit-oscillator coherences, the core signature of high-dimensional correlation}. We 
demonstrate {characteristic} visibilities of $20\%$ and $40\%$,  which approach the theoretical visibility limit, providing a scalable coherence witness {without} full state tomography for high-dimensional states. 
%Reaching this ceiling is a direct confirmation of the protocol's high fidelity in generating the target coherent state.
%This is confirmed by a interference contrast and visibility of approximately $40\%$,  a value limited not by decoherence but by the measurement scheme itself, which matches the ideal theoretical prediction.}

%—a direct signature of the balanced interference pathways from states with a mean motional number exceeding four quanta and a {similar standard deviation}.
 %, we directly observe a substantial interference visibility of approximately $40\%$ for entangled states with a mean motional number exceeding four quanta,  and a {similar standard deviation}.   
%{The scaling of this contrast visibility with the state's complexity closely tracks ideal theoretical predictions.}
% This result {establishes a proof-of-principle}  protocol {resilient against}  {a stable but unknown interaction phase}, %passively resilient to common experimental imperfections, relaxing the need for active phase stabilization and 
% {and provides a design principle for  passively stabilized generation of complex quantum states}.

\end{abstract}

\maketitle

%\section{Introduction}

%\section{Introduction}

%The superposition principle, a foundational tenet of quantum mechanics, posits that systems can exist in multiple states simultaneously, giving rise to interference phenomena even for objects of considerable scale and complexity \cite{Leggett1980, Arndt1999}. Generating and verifying such interference for ``macroscopic'' states—those characterized by large phase-space separations or high excitation numbers—is paramount for fundamental tests of quantum theory and for the advancement of quantum technologies \cite{Frowis2018RevModPhysMacroscopic,KovachyNature2015SuperposeCenti, ClarkeQST2019Macroscopic, LoNature2015SpinmotionSqz, Millen2020, BrunelliPRA2018ReservoirEngineering, CardosoPRA2021TMSV, Vlastakis2013SuperconductingCat, Aspelmeyer2014OptomechanicsReview}. Trapped ions, offering exquisite control over both internal and motional degrees of freedom, serve as an exemplary platform for these sophisticated explorations \cite{Leibfried2003, JarlaudJPB2021CoherenceMotional, OhiraPRA2022CoherentCononDynamicDecoupling, WeidtPRL2015CoolingRadiation}.
%Mutual interference between quantum systems is a fundamental effect in quantum mechanics and a pivotal element of modern quantum technologies. 
Quantum interference  between interacting systems is fundamental to quantum technology.
Open systems naturally suffer from decoherence, which limits quantum interference \cite{ZurekRevModPhys2003Decoherence}. 
For nearly closed systems approaching unitary dynamics and initial states close to pure, the interference effect can be  exquisitely sensitive to  control imperfections.
In many systems, such control is provided by a laser (or microwave field) through its phase. 
The nature of the laser phase itself has been {thoroughly studied}~\cite{MolmerPRA1997Coherence,Wiseman10.1117/12.497090, RudolphPRL2001CoherenceTeleportation, VanEnkPRL2002Laser,Ludlow2015RevModPhys.87.637AtomicClocks}. 
%One perspective argues that a laser's output is not a single coherent state but a statistical mixture over all possible phases. 
If such a phase fluctuates rapidly, this acts as stochastic noise, quickly destroying coherence and resulting in a fully mixed state. 
% Our work, however, addresses a different, common experimental challenge: a phase that is \textit{stable} during an experimental run but has an unknown, uncontrolled value. 
% This represents a static imperfection rather than a dynamic noise source, and it is this robustness that our protocol demonstrates. 
%The most provocative and discussed examples are Schr\"odinger cat states \cite{Schrodinger1935, HarocheRaimond2006}, where quantum systems with low-dimensional Hilbert space (like a qubit) interfere with systems having a high-dimensional one (like an oscillator). 
However, does limited knowledge of a stable phase preclude complex quantum interference experiments?

A particularly challenging quantum scenario involves interference between a low-dimensional system (a qubit) and a high-dimensional one (an oscillator), forming states analogous to Schrödinger's cat \cite{Schrodinger1935, HarocheRaimond2006, Jeong:AmplifyingHybrid}. 
% Many experiments over different platforms have already demonstrated such cat states \cite{Monroe1996, Brune1996, Vlastakis2013}, analyzed their features \cite{Leibfried2005, Frowis2018}, and used their interference aspects to build advanced sensing \cite{Joo2011, McCormick2019} and error correction protocols \cite{Leghtas2013, Ofek2016}. 
% However, the interaction that builds such  states towards the macroscopic limit, while being close to unitary, {is highly likely to} have parameters that are stable but not precisely known.
% Such imperfect control due to such a partial knowledge {of the parameters} critically prohibits observation of interference, {making experiments targeting the macroscopic limit extremely difficult.}
While such states have been demonstrated across various platforms \cite{Monroe1996, Brune1996, Vlastakis2013,Leibfried2005, Frowis2018} and are crucial for sensing and bosonic quantum error correction, where information is stored in the parity of the oscillator \cite{GottesmanPRA2001GKP,Joo2011, McCormick2019, Leghtas2013, Ofek2016, Fluehmann2019}, their creation can be {extremely} sensitive to interaction phase \cite{HempelNatPho2013}, relying on complex active feedback or precise phase stabilization \cite{HempelNatPho2013, KienzlerScience2015,Gerlich2011}.
A primary obstacle is precisely knowing and controlling the phases of the interactions that are linked to the laser phase. 
% If the interaction phase is completely random between each run, averaging thousands
% of measurements mixes constructive and destructive interference.
% This critically prohibits a cat state observation, as the averaging washes out the interference. 
{If the interaction phase is random between runs, averaging measurements mixes constructive and destructive interference, precluding cat state observation.}
% fringes, whose contrast is the direct measure of underlying quantum coherence.
In many realistic situations, such a laser phase can be stable (non-stochastic) during an experimental run over the timescale of a single experimental sequence ($\mu$s) but not perfectly known or controlled {between sequences}.
%Because this phase varies randomly from one run to the next, averaging the thousands of runs required for statistics would cause peaks of one interference pattern to land in the troughs of another, causing the final averaged signal for a conventional protocol to wash out completely.
%{If the phase is random between each run, averaging thousands of measurements mixes constructive and destructive interference, causing the overall signal to average to zero. }
%This critically prohibits its observation. 
%The conventional approach is to eliminate such errors through resource-intensive active stabilization, that hinders scalability.

In this work, we address such a situation by experimentally generating a quantum interference between a trapped-ion qubit and its motion using a preparation sequence of alternating blue and red sideband pulses on a trapped-ion's oscillator. 
This sequence deterministically enforces a \textit{qubit energy and oscillator number-parity  correlation {and their} interference}. %, where the qubit’s excited state is exclusively paired with an odd number of  oscillator quanta, and its ground state with an even number. 
%This parity structure, {a structural invariant of the alternating pulse sequence, is inherently} compatible with binomial quantum error correction codes \cite{MichaelPRX2016}, making our method {compatible with} for logical qubit preparation. 
{This qubit-parity structure aligns with cat codes \cite{CochranePhysRevA.59.2631,RalphPhysRevA2023.68.042319,mirrahimi2014dynamically} and general binomial quantum error correction codes \cite{MichaelPRX2016}, facilitating logical parity manifold preparation.}
%{This underlying structure makes the state's logical properties immune to unknown phases in the control field.} 
%{This makes the global state structure a topological invariant of the sequence, ensuring passive resilience against laser phase shifts.}
Crucially, while a stable but unknown phase in the laser drive can alter the specific amplitudes \textit{within} the oscillator state, the  underlying qubit-parity correlation {and interference} remains locked by the symmetry of the sideband interactions. 
%This intrinsic robustness allows us to achieve $40\%$  visibility in interference between the qubit and an oscillator state with an increasing mean phonon number. 
{To characterize this qubit-parity interference, we measure two minimal visibilities: internal oscillator coherence and  mutual qubit-oscillator interference. }
This {structural invariance} allows us to achieve $40\%$ visibility in internal oscillator interference and $20\%$ visibility  in the qubit-oscillator interference with an increasing mean phonon number, {up to} $\av{\hat{n}}\approx4$, with standard deviation $\Delta \hat{n} \approx 4$, reproducing the theoretical prediction {in a scalable way}.
%{This protocol creates a quantum state with qubit-parity structure within an engineered, symmetric noiseless subsystem with respect to the unknown interaction phase, showcasing a passive resilience} {to this specific control error. The protocol remains sensitive to general decoherence, such as motional heating and spontaneous emission, which becomes the primary limitation for longer pulse sequences.}
%We observe over 40\%, despite the unknown phase in the cat state generation process. 
%We use a motional single-trapped-ion oscillator coupled to electronic levels and controlled for this demonstration. 

For this demonstration, we use a single $^{40}$Ca$^{+}$ ion in a linear Paul trap. 
The qubit is encoded in the long-lived ground and excited  Zeeman sublevels, and the oscillator is a motional mode of the ion.
%The qubit is encoded in the long-lived $|4^2S_{1/2}, m = -1/2\rangle$ (ground) and $|3^2D_{5/2}, m = -5/2\rangle$ (excited) states, and the oscillator is a motional mode of the ion.}
The optical controls we employ induce elementary Jaynes-Cummings (JC) {and anti-JC} interactions  \cite{Leibfried2003}, which controllably entangle the qubit and motional states with an unknown {but stable (repeatable)} phase. 
This method is technically similar to the schemes that employ specific sideband structures for complex state generation \cite{Alderete2021, Matsukevich:BosonicSculpting}, {and builds upon deterministic, measurement-free preparation of logical states \cite{Hastrup2021}}.
%However, the JC interactions crucially depend on the internal phase connected to optical pulse controls, which has a stable but uncontrolled value  in our proof-of-principle experiment. 
%To observe the qubit-oscillator interference, we use a simplified detection sequence of one red and blue pulse with two controllable relative phases that are evaluated for the interference fringes. 
To analyze the interference, we employ a two-stage verification process: a single-pulse scan first confirms the presence of basic quantum interference between qubit and oscillator,  {but is insensitive to} internal oscillator coherence. 
Subsequently, a two-pulse sequence with two controllable relative phases {minimally but} unambiguously resolves the distinct contributions from both qubit-oscillator and internal oscillator coherence. 
%Our study complements the previous ideas on the hot cat states \cite{Zhu1996Quantum, Huyet2001Superposition, Jeong2006Transfer, Zheng2007Macroscopic, Jeong2007Quantum, Nicacio2010Phase}, culminating with a recent experiment \cite{Yang2025SciAdv2025HotCat} where interference has been finally observed despite the initial thermal oscillator noise.
Our study complements  a  recent experiment on `hot' cat states \cite{Yang2025SciAdv2025HotCat}. 
While those experiments demonstrate that quantum coherence can be generated \textit{from} an imperfect, thermal initial state, we address a complementary concept: generating %coherence \textit{despite} an imperfect, uncontrolled interaction phase during the state preparation dynamics, 
and  verifying qubit-oscillator interference despite imperfect, uncontrolled interaction dynamics.

\begin{figure}
 \includegraphics[width=1.0\textwidth]{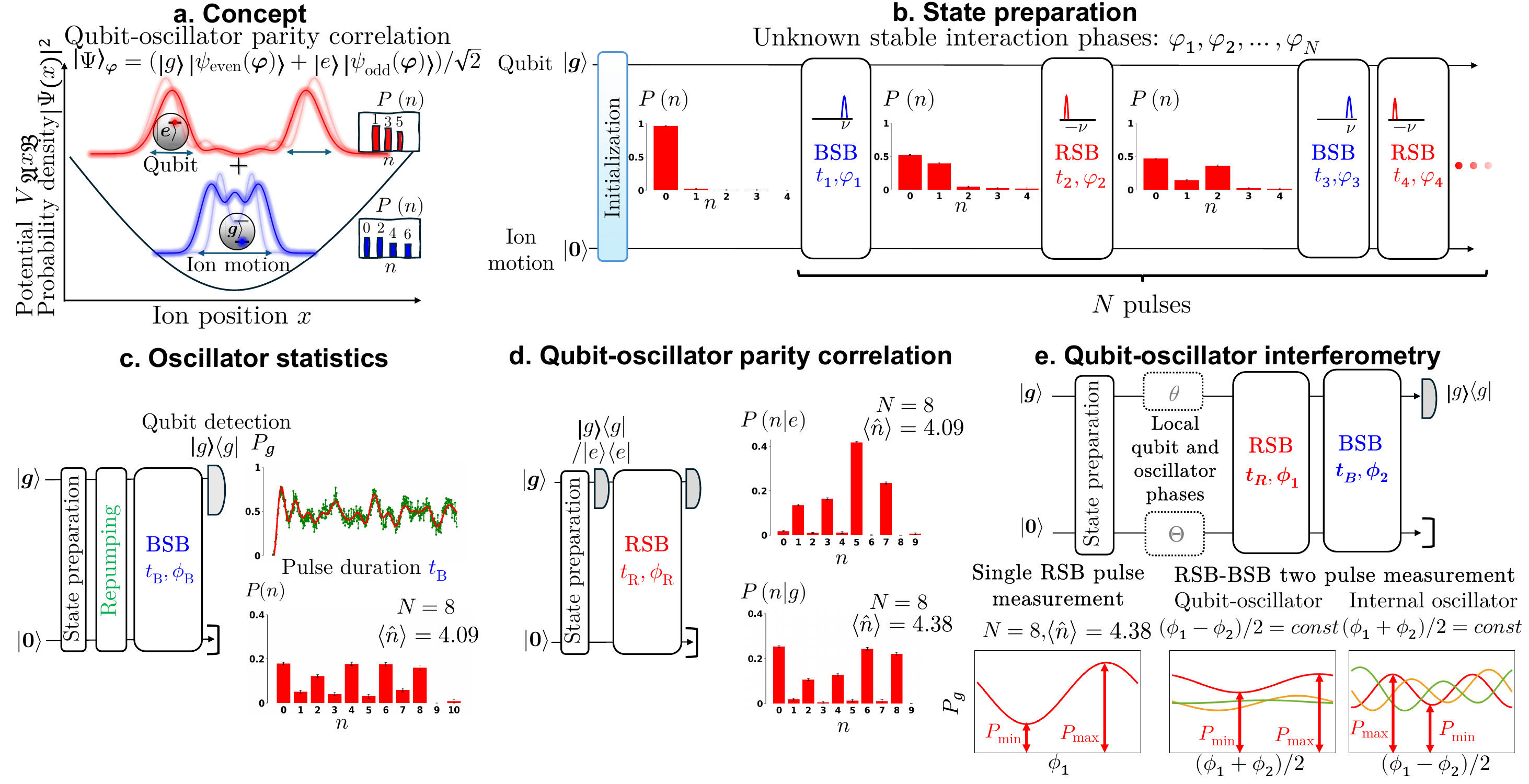}
 \caption{
 %Protocol concept. It is made of two stages: state preparation, and coherence verification. {All pulse coupling parameters such as pulse lengths ($t$), Rabi frequencies ($\Omega_0$), Lamb-Dicke parameter ($\eta$), are kept constant. All the preparation pulses phases $\phi_j$ are not controlled. The verification can be included in the preparation, where the phase of the last two pulses of the preparation can be varied for verification.
% Protocol Concept: State Preparation and coherence Verification.
% All pulse parameters—pulse lengths $t$, fundamental Rabi frequencies $\Omega_0$, and Lamb-Dicke parameter $\eta$—are kept constant. The phases $\phi_j$ of the preparation pulses are uncontrolled. \textbf{a} Mutual coherence verification requires varying one verification pulse phase, using their equivalence in physical effect. \textbf{b} Intra coherence verification takes place after the preparation by varying the phases of the two verification pulses and detecting the qubit.
%Overview of the protocol for creating and verifying  and qubit-parity non-local coherence with macroscopic motional superpositions in a trapped ion. 
\textbf{Creating and verifying  qubit-oscillator {interference} and internal oscillator coherence in a trapped ion {despite unknown interaction phases}. }
\textbf{a} Visualization of the  qubit-parity correlation in the state $\ket{\Psi}_{\bm{\varphi}}$ (\ref{eq:psi}), where the qubit state dictates the parity of the motional state. 
%Qubit-parity {correlation} is the key {property}, leading to spatially distinct motional states $\ket{\psi_\pm}$ with odd and even parity {depending on the qubit energy eigenstates} $\ket{g(e)}$, visualized above.
\textbf{b}  The state preparation sequence, consisting of alternating blue- and red-sideband (BSB, RSB) pulses producing Jaynes-Cummings and anti-Jaynes-Cummings interactions with stable but unknown interaction phases $\varphi_i$.
 \textbf{c} The measured phonon number distribution $P(n)$ obtained via Rabi flopping (described in Appendix \ref{appendix:experiment}), showing a significant mean phonon number ($\av{\hat{n}}=4.09$) and standard deviation. %(Figure \ref{fig:new_fig2}\textbf{a,b}).
 The BSB pulse duration {$t_\mathrm{B}$} is varied, while the pulse phase {$\phi_\mathrm{B}$ is fixed.}
\textbf{d} Measurement of the qubit-parity correlation, confirming that the ground (excited) qubit state is associated with even (odd) phonon numbers (Figure \ref{fig:new_fig2}\textbf{c,d}).
The RSB pulse duration {$t_\mathrm{R}$} is varied, while the pulse phase {$\phi_\mathrm{R}$} is fixed.
\textbf{e} A two-pulse verification sequence with experimentally scanned phases $\phi_1$ and $\phi_2$ used to measure interference fringes in the final qubit ground state population $P_g$, from the  state in \textbf{c} and \textbf{d} {($N=8, \langle \hat{n} \rangle \approx 4$)}. {Solid lines represent fits to experimental data; different colors indicate distinct phase constraints.} %having roughly the same average phonon number $\av{\hat{n}}$. 
Without full state tomography and active control of local phases $\theta$ and $\Theta$, this measurement directly distinguishes between the two fundamental types of coherence:
\textit{Qubit-oscillator interference} (locally {controllable by the qubit phase} $\theta$) represents the coherence between the qubit's state and the motion's parity.  The lowest order qubit-oscillator interference can be  sufficiently measured using a single red pulse measurement.  
Its higher order interference is probed by scanning the phase sum of the two verification pulses, $\frac{\phi_1 + \phi_2}{2}$ while the phase difference $\frac{\phi_1 - \phi_2}{2}$ is constrained by different values.
\textit{Internal oscillator coherence} (locally {controllable by the oscillator phase} $\Theta$) representing coherence between different motional components within a single parity subspace, is probed by scanning the phase difference $\frac{\phi_1 - \phi_2}{2}$, while the phase sum is held constant by different values.
The contrast and visibility of all these curves are averaged over the phases $\varphi_i$ for $i=1,...,N$. 
}
\label{fig:concept}
\end{figure}

We consider the generation of a balanced, qubit-oscillator superposition state of the form, {depicted in Fig. \ref{fig:concept}\textbf{a}}.
\begin{align}
\ket{\Psi}_{\bm{\varphi}}=2^{-1/2}\left(|e\rangle|\psi_{\text{odd}}(\bm{\varphi})\rangle + |g\rangle|\psi_{\text{even}}(\bm{\varphi})\rangle\right).
\label{eq:psi}
\end{align}
Here, $\ket{g}$ and $\ket{e}$ are the qubit energy eigenstates. 
The vector of $N$ preparation parameters is defined as $\bm{\varphi} = (\varphi_1,\varphi_2,\dots,\varphi_N) \in \mathbb{R}^N$ \cmmnt{represents the unknown stable phases for the $N$ preparation pulses}.
Each $\varphi_j \in [0, 2\pi)$ is a static offset that remains constant throughout a single experimental sequence but varies between realizations {of such sequences}, effectively behaving as a systematic parameter rather than stochastic noise.
%is the ordered set of all $N$ individual {interaction} phases used in the state preparation sequence {of $N$ pulses}, assumed to be stable but unknown. 
In such a state, $|\psi_{\rm odd}(\bm{\varphi})\rangle = \sum_{n=\text{odd}} c_n (\bm{\varphi}) |n\rangle$ and $|\psi_{\rm even}(\bm{\varphi})\rangle = \sum_{n=\text{even}} c_n (\bm{\varphi}) |n\rangle$ {with $c_n\in \mathbb{C}$} are orthogonal motional  {eigenstates with opposite parity}, $\langle(-1)^{\hat{n}}\rangle = \pm1$, respectively. Here, $\ket{n}$ is an oscillator Fock state.
This state $\ket{\Psi}_{\bm{\varphi}}$  in (\ref{eq:psi}) is characterized {minimally} by two distinct phase-sensitive effects. First, the state possesses \textit{mutual qubit-{parity} interference}, representing a stable phase relationship between the qubit state and the oscillator's parity. This coherence leads to interference effects that can be controlled by the relative phase $\theta$:
\begin{equation}
    |\Psi\rangle_{\theta} = 2^{-1/2} \left(e^{\ii\theta}|e\rangle |\psi_{\rm odd}(\bm{\varphi})\rangle + |g\rangle |\psi_{\rm even}(\bm{\varphi})\rangle\right).
\end{equation}
Here, phase $\theta$ can be controlled by a local qubit phase rotation by a unitary operator $e^{\ii\theta \hat{\sigma}_z/2}$, up to a global phase.

Second, the motional states  {$|\psi_{\rm odd/even}(\bm{\varphi})\rangle$} themselves possess \textit{internal oscillator coherence}, {a stable phase relationship among their constituent Fock states. This coherence can be {analogously}  probed by applying a local oscillator phase} $\Theta$. This is described by a motional phase rotation, where the operator $\exp[\ii\Theta\hat{n}]$ applies a phase $e^{\ii\Theta n}$ to each Fock state component $|n\rangle$. The complete state, including both types of phase-sensitive effects, is then given by:
\begin{equation}
    |\Psi\rangle_{\theta,\Theta} = \exp[\ii\Theta\hat{n}]|\Psi\rangle_{\theta} = 2^{-1/2}\left(e^{\ii\theta}|e\rangle|\psi_{\rm odd}(\bm{\varphi}, \Theta)\rangle + |g\rangle|\psi_{\rm even}(\bm{\varphi}, \Theta)\rangle\right),
    \label{eq:psiTheta}
\end{equation}
where the motional states are defined as $|\psi_{\rm odd/even}(\bm{\varphi}, \Theta)\rangle \equiv \exp[\ii\Theta\hat{n}]|\psi_{\rm odd/even}(\bm{\varphi})\rangle$.

% The parameter $\theta$ quantifies the mutual qubit-oscillator coherence, while $\Theta$ parameterizes the internal coherence of the oscillator states. These two effects are distinct:
%   (i) The \textit{mutual qubit-oscillator coherence} is the single relative phase, $\theta$, between the two orthogonal qubit-parity subspaces, $|e\rangle|\psi_{\text{odd}}\rangle$ and $|g\rangle|\psi_{\text{even}}\rangle$.
%     (ii) The \textit{internal oscillator coherence} refers to the stable phase relationships \textit{within} each motional state, e.g., between the $|0\rangle$ and $|2\rangle$ components of $|\psi_{\text{even}}\rangle$. This complex phase structure is controlled by the parameter $\Theta$.
    
%A key aspect of our work is experimentally distinguishing and {demonstrating the effects of} these two coherences independently. 
A two-pulse {measurement} sequence  provides an orthogonal mapping of the qubit-oscillator and internal oscillator coherences, characterizing their respective contributions to the observed interference, thus validates the state structure.
The core principle of {robust qubit-oscillator interference and internal oscillator coherence is} this qubit-parity {correlation in the} superposition state $|\Psi \rangle_{\theta,\Theta}$, preserving the orthogonality of the motional {parity} subspaces regardless of the values in $\bm{\varphi}$.

The phase $\theta$ is the relative phase between the two components of the superposition in Eq. (\ref{eq:psi}). 
It represents the coherence between the qubit being in $|g\rangle$ with an even-parity motion and in $|e\rangle$ with an odd-parity motion.
% If this phase $\theta$ were to become unstable between experimental runs, this coherence and the resulting qubit-oscillator interference would be lost. 
% The system would then be described by the mixed state by qubit dephasing:
{Instability in $\theta$ between runs destroys this coherence, reducing the system to a mixed state described by qubit dephasing:}
\begin{align}
    \rho_\mathrm{Q-dep} = \tfrac{1}{2} \ket{e}\bra{e} \otimes \ket{\psi_-}\bra{\psi_-} + \tfrac{1}{2} \ket{g}\bra{g} \otimes\ket{\psi_+}\bra{\psi_+}.
    \label{eq:incoherentDM2}
\end{align}
Here, the motional states $\ket{\psi_\pm}$ retain their internal coherence but are only classically correlated with the qubit state. 
If strong motional dephasing also occurs, or if all the preparation pulses are unstable, this {motional} coherence is also lost, resulting in a state {fully} diagonal in the {qubit-Fock} basis:
\begin{align}
    \rho_\mathrm{O-dep} = \tfrac{1}{2} \ket{e}\bra{e} \otimes \mathrm{diag}[\ket{\psi_-}\bra{\psi_-}] + \tfrac{1}{2} \ket{g}\bra{g} \otimes \mathrm{diag}[\ket{\psi_+}\bra{\psi_+}].
    \label{eq:incoherentDM}
\end{align}
Our experiment aims to demonstrate both qubit-oscillator interference and internal oscillator coherence despite stable but unknown phases {of the preparation pulses}, distinguishing the pure state $\ket{\Psi}$ from these incoherent mixtures ($\rho_\mathrm{Q-dep}$, $\rho_\mathrm{O-dep} $), as detailed in Appendix~\ref{Appendix:decoherence}.

%\subsection*{Qubit-parity interference}

\subparagraph{Preparation}
%Below,  we denote states in terms of  $\ket{g (e),n}$ in a qubit-Fock joint basis, where $\ket{g(e)}$ are qubit energy eigenstates, and  $\ket{n}$ is the oscillator Fock state. 
%We use the qubit-Fock basis $\ket{g(e), n}$, where $\ket{g}$ and $\ket{e}$ are the qubit energy eigenstates and $\ket{n}$ is an oscillator Fock state. %{which does not show any Ramsey-type modulation}.
We describe the states in the qubit-Fock basis $\ket{g(e), n}$. We begin {with} the qubit-oscillator ground state $|g,0\rangle$ achieved by cooling, and apply a sequence of alternating blue sideband (BSB) and red sideband (RSB) pulses for preparation depicted in Fig. \ref{fig:concept}\textbf{b}, {which are} defined with details in Appendix \ref{Appendix:pulse}. 
%These pulses are timed and phased to produce the states with large average phonon number and its standard deviation, also advantageous for subsequent coherence verification. 
BSB (RSB) pulses increase the maximum motional quantum number while exciting {(de-exciting)} the qubit, as well as the standard deviation of the Fock distribution {for a broad range of pulse areas}.
%; RSB pulses decrease it, de-exciting the qubit. 
Together, these alternating pulses coherently drive the system to the desired qubit-motional state.
The total {$N$} pulse sequence is described by {the unitary} operator $\hat{U}_\mathrm{prep}$, a sequential product of $N$ anti-Jaynes-Cummings (BSB, $\hat{U}_{\text{aJC}}$) and Jaynes-Cummings (RSB, $\hat{U}_{\text{JC}}$) interactions as in Fig. \ref{fig:concept}\textbf{b}:%Jaynes-Cummings ($\hat{U}_\mathrm{JC}{(j)}'$) and anti-Jaynes-Cummings ($\hat{U}_\mathrm{aJC}{(j)}$) interactions, for j=1 to j=N/2: 
\begin{align}
    \hat{U}_\mathrm{prep}=\prod_{j=1}^{N/2} \hat{U}_\mathrm{JC}[t_{2j},\varphi_{2j}]\hat{U}_\mathrm{aJC}[t_{2j-1},\varphi_{2j-1}],%\equiv \prod_{j=1}^{N/2} \hat{U}_\mathrm{JC}^{(2j)}\hat{U}_\mathrm{aJC}^{(2j-1)}.
    \label{eq:prep}
\end{align}
where $\varphi_{j}$ comprise $\bm{\varphi}$ in Eq. (\ref{eq:psi}).
{For odd $N$, the preparation sequence ends with BSB pulse, without the final RSB pulse.}

In our experiment, we relax the  challenging constraint on small time steps of $t_j \ll 1$ and perfectly controlled phases of $\varphi_j$, typically needed for  Suzuki-Trotter decomposition \cite{MezzacapoSREP2014DigitalSimulationRabi}. 
Our approach instead accommodates \textit{stable but unknown} phase offsets in the {phases of laser} driving fields ($\varphi_j$ in Eq.~\eqref{eq:prep}) inherent to the setup (e.g., {corresponding to the stable relative phases within the common laser-qubit coherence and assuming the motional coherence to be much longer})  throughout the experiment but are not actively controlled  to specific target values. 
Quantitatively, the phase remains constant over the timescale of a single experimental sequence (on the order of {milliseconds}), with {particular limits on the} coherence time detailed in Appendix \ref{appendix:experiment}.
These stable but unknown phases have a crucial distinction from stochastic, shot-to-shot noise of these interaction phases $\varphi_j$, to which this {method} is not immune \cite{Wiseman:NonMarkovianity}.
% {It is crucial to distinguish the phase error our protocol addresses from other forms of noise. 
% We demonstrate immunity to an interaction phase $\varphi_j$ of the driving laser that is stable within a single run but varies randomly shot-to-shot. 
% Our protocol is not, however, designed to be immune to random fluctuations in the relative phases between consecutive pulses within a single sequence, which would constitute a different form of decoherence. }
%a crucial distinction from random, shot-to-shot phase noise. 
% This approach significantly simplifies the experiment by avoiding demanding active phase stabilization techniques. 
% The protocol's robustness to these unknown-but-stable phases stems directly from the structured sequence of alternating red and blue sideband pulses for the preparation of the qubit-parity entangled form.
%These pulses are timed, but phase is uncontrolled, creating states with high average phonon number and variance, beneficial for coherence verification.
%About the effect of the applied pulses, see Appendix \ref{Appendix:pulse}.

%To initiate an experimental run, we explicitly consider the  first two preparation pulses 
{The  first preparation pulse acts on the initial state} $|g, 0\rangle$, where the qubit is in ground state $|g\rangle$ and the motional state has an \textit{even} phonon number ($n=0$).
{The second pulse already acts on the qubit-oscillator correlation.}
Following Fig.~\ref{fig:concept}\textbf{b} we apply:
\begin{enumerate}
    \item \textbf{First (BSB) pulse:} A blue-sideband pulse can only transition the system to $|e, 1\rangle$ {as $|\Psi_1\rangle = \cos(t_1)\ket{g, 0} + e^{\ii\varphi_1}\sin(t_1)\ket{e, 1}$}. 
The qubit state $|e\rangle$ is now linked exclusively to an \textit{odd} phonon number ($n=1$).
    \item \textbf{Second (RSB) pulse:} A subsequent red-sideband pulse can only act on the new $|e, 1\rangle$ component, transitioning it back to $|g, 2\rangle$ {as $|\Psi_2\rangle = \cos(t_1)\ket{g, 0} + e^{\ii\varphi_1}\sin(t_1)\cos(t_2)\ket{e, 1} + e^{\ii(\varphi_1+\varphi_2)}\sin(t_1)\sin(t_2)\ket{g, 2}$}.
The qubit state $|g\rangle$ is thus again linked only to an \textit{even} phonon number.
\end{enumerate}
This alternating sequence, that can be extended further,  enforces the qubit-parity correlation throughout the circuit. 
The \cmmnt{unknown-but-stable} phases $\varphi_j$ \cmmnt{of the preparation pulses only determine the complex amplitudes of these transitions, not the fundamental qubit-parity correlation structure of the resulting state.} {modulate transition amplitudes but leave the qubit-parity structure invariant.}

The alternating sideband pulses described in Fig. \ref{fig:concept}\textbf{b} enforce a strict qubit-parity correlation---linking the qubit state $\ket{g}$ to even phonon numbers $n$ and $\ket{e}$ to odd $n$---which ensures the generation of the target state structure (Eq.~(\ref{eq:psi})) regardless of the phases $\bm{\varphi}$. 
\cmmnt{Ideally, the experiment described in Fig. \ref{fig:concept}\textbf{b} generates a state with the desired qubit-parity {correlated structure} $\ket{\Psi}_{\theta,\Theta}$.}
% :
% \begin{align}
% \ket{\Psi} = e^{\ii\theta}\ket{e}\sum_{n=0} c_{2n+1}(\bm{\varphi})\ket{2n+1} + \ket{g}\sum_{n=0} c_{2n}(\bm{\varphi})\ket{2n} \equiv e^{\ii\theta}\ket{e}\ket{\psi_\mathrm{odd}(\bm{\varphi})} + \ket{g}\ket{\psi_\mathrm{even}(\bm{\varphi})}.
% \label{eq:globalcoherent}
% \end{align}
% Here, $|\psi_\mathrm{odd(even)}(\bm{\varphi})\rangle$ {in (\ref{eq:globalcoherent})} are motional states with exclusively odd (even) phonon numbers. 
While this fundamental structure is preserved, the specific complex coefficients $c_n(\bm{\varphi})$ and the overall relative phase $\theta$ depend on the \cmmnt{uncontrolled unknown but fixed} preparation phases and pulse lengths. 
To simplify the experimental demonstration, the target pulse area for each step in our experiment is set to $\eta\Omega_0t_j \approx \pi/4$ (i.e. $\pi/2$ pulse), which corresponds to a half-transfer of population for the $n=0 \leftrightarrow n=1$ transition.

Following the state preparation sequence, we characterize the resulting state and verify its fundamental structure. 
First, we measure the phonon number distribution $P(n)$, confirming the creation of a macroscopic motional state with a significant mean phonon number, as conceptually depicted in Fig.~\ref{fig:concept}\textbf{c}. 
Subsequently, we verify the state's key structural feature by measuring the qubit-parity correlation (Fig.~\ref{fig:concept}\textbf{d}). 
This confirms that the qubit ground state $\ket{g}$ is exclusively associated with even phonon numbers, while the excited state $\ket{e}$ is associated with odd phonon numbers, a correlation central to our {experimental} robustness.
{The details of these experiments are summarized in Appendix \ref{appendix:experiment}.}

\subparagraph{Interference by single-pulse measurement}

Our first step is to distinguish the prepared state $\ket{\Psi}$ from a fully incoherent mixed state, such as $\rho_{\text{O-dep}}$ (Eq.~(\ref{eq:incoherentDM})). We apply a single RSB verification pulse with a variable phase $\phi_1$, which drives transitions like $\ket{g, 2n} \leftrightarrow \ket{e, 2n - 1}$. Scanning the phase $\phi_1$ of this single verification pulse directly creates  and reveals interference between the $|g\rangle|\psi_\text{even}(\bm{\varphi})\rangle$ and $|e\rangle|\psi_\text{odd}(\bm{\varphi})\rangle$ components of the state~(\ref{eq:psi}). \cmmnt{This directly creates interference between the even-parity $\ket{g}\ket{\psi_{\text{even}}}$ and odd-parity $\ket{e}\ket{\psi_{\text{odd}}}$ components of the state~(\ref{eq:psi}). }
If the state possesses qubit-oscillator coherence {in subspaces $\{\ket{g,2n}, \ket{e,2n-1}\}$ for any integer $n$}, the final qubit population $P_g$ will modulate as a sinusoidal function of $\phi_1$ {(Fig.~\ref{fig:concept}\textbf{e}, left)}. %To quantitatively distinguish our prepared state from an incoherent mixture, we model this interference using a single coherence factor, $w$, as detailed in Appendix~\ref{Appendix:decoherence}.
%However, this single-pulse method is limited: the observed modulation confirms the \textit{presence} of first order qubit-oscillator coherence, but as shown in Appendix \ref{Appendix:decoherence}, it is insensitive to the internal oscillator coherence within each qubit subspace of the prepared states, and blind to the higher order qubit-oscillator coherence.
While this confirms  {some interference}, it cannot resolve internal oscillator or {coherence}, necessitating a longer sequence.
The prepared state may still have only qubit-oscillator coherence, in the {density matrix of $\rho=\sum_n \ket{\psi_n}\bra{\psi_n}$, where $\ket{\psi_n}=c_{2n}\ket{g,2n}+c_{2n-1}\ket{e,2n-1}$} with complex coefficients $c_j$ compatible with the observed qubit-oscillator parity correlation. 
Therefore, the single-pulse measurement acts as a crucial witness of coherence, confirming the prepared state is not a simple incoherent mixture {such as $\rho_\mathrm{Q-dep}$ (\ref{eq:incoherentDM2}) or $\rho_\mathrm{O-dep}$ (\ref{eq:incoherentDM}),  while it cannot on its own unambiguously resolve them}.
{To systematically characterize the higher-order coherence contributions probed in our measurements}, throughout the work, we define the $k$-th order off-diagonal 
Fock elements as $|\langle n | \rho_O | n-k \rangle|$ for any integer $n$ {with $0\le k\le n$}, where $\rho_O$ is the reduced density matrix of the oscillator.
Due to the choice of verification pulse length $t_\mathrm{R}=\pi/2$, the dominant term contributing to the contrast {and visibility here} is from $\bra{2}\rho_O\ket{1}$, {which belongs to the $1$st order off-diagonal}.

\subparagraph{Interference by two-pulse measurement}

We next apply  a two-pulse verification sequence consisting of a RSB pulse with phase $\phi_1$ followed by a BSB pulse with phase $\phi_2${, as in Figure \ref{fig:concept}\textbf{e}}. 
%A crucial aspect of this method is the {orthogonal}  mapping between the experimental control parameters---the verification pulse phases $\phi_1$ and $\phi_2$---and the {local} phases $\theta$ and $\Theta$ of the prepared state $\ket{\Psi}_{\theta, \Theta}$ in (\ref{eq:psiTheta}). 
This method maps the experimental verification phases 
$\phi_1$ and $\phi_2$  to the local  phases $\theta$ and $\Theta$ (Eq. \ref{eq:psiTheta}).
This relationship is  a direct consequence of the measurement interaction, as we derive from the operator commutation relations in Appendix \ref{appendix:equivalnece}. 
%The derivation shows that the final qubit population $P_g$ depends on the difference between the state's internal coherence phases and specific combinations of the pulse phases. 
The two verification pulses  project the prepared state onto a measurement basis of the form
\begin{align}
    \ket{\Psi_m(\phi_1, \phi_2)}=2^{-1/2}\left(|e\rangle|\psi^\text{(measure)}_\mathrm{odd}(\phi_1, \phi_2)\rangle+|g\rangle|\psi^\text{(measure)}_\mathrm{even}(\phi_1, \phi_2)\rangle\right),
    \label{eq:psim}
\end{align}
%where %$\theta_{\text{det}}$ and $\Theta_\mathrm{det}$ are a combination of the externally controlled phases of the verification pulses (e.g., $\phi_1, \phi_2$ in Fig.~1e).  
%$(\phi_1, \phi_2)$ for two pulse verification schemes  is the vector of   individual pulse phases used in the measurement sequence. %are defined similarly as $\bm{\varphi}$.
%The resulting interference depends on the phase differences between the prepared state and the measurement basis, i.e., on $\theta - \theta_{\text{det}}$ and $\Theta - \Theta_{\text{det}}$.
where $|\psi^\text{(measure)}_\mathrm{odd}(\phi_1, \phi_2)\rangle$ and $|\psi^\text{(measure)}_\mathrm{even}(\phi_1, \phi_2)\rangle$ are also distinct, orthogonal motional states of the oscillator that possess opposite parity, but may have different (here typically smaller) average number of quanta than $|\psi_\mathrm{odd}(\bm{\varphi})\rangle$ and $|\psi_\mathrm{even}(\bm{\varphi})\rangle$. 
We then measure the final ground state population, 
\begin{align}
P_g=|\braketk{\Psi_m(\phi_1,\phi_2)}{\Psi}_{\bm{\varphi}}|^2= \frac{1}{4} | \langle \psi_{\text{even}}^{\text{(measure)}} | \psi_{\text{even}}(\Theta) \rangle + e^{\ii\theta} \langle \psi_{\text{odd}}^{\text{(measure)}} | \psi_{\text{odd}}(\Theta) \rangle |^2,
\end{align}
{that oscillates as a function of the phases $\phi_{1,2}$.}
%\cmmnt{Scanning the phase $\phi_1$ of a single verification pulse reveals interference between the $|g\rangle|\psi_\text{even}(\bm{\varphi})\rangle$ and $|e\rangle|\psi_\text{odd}(\bm{\varphi})\rangle$ components.} %, {respective to a phase} {$\theta$ and $\Theta$, indistinguishably}.   
{Due to the choice of verification pulse length $t_\mathrm{R}=\pi/2$, the dominant term contributing to the contrast by this measurement is from $\bra{2}\rho_O\ket{1}$ and $\bra{4}\rho_O\ket{1}$. }

Specifically, as proven by the unitary transformations in Eqs. (E1–E4), the final qubit population $P_g$ depends functionally on the terms $\theta-\frac{\phi_1+\phi_2}{2}$ {(qubit-oscillator interference)} and $\Theta-\frac{\phi_1-\phi_2}{2}$ {(internal oscillator coherence)}, and thus scanning the phase sum $(\phi_1 + \phi_2)/2$  has the identical effect on the modulation by the qubit-oscillator coherence $\theta$ for fixed phase difference $(\phi_1 - \phi_2)/2$, while scanning the phase difference $(\phi_1 - \phi_2)/2$ has the identical  modulation by the internal oscillator coherence $\Theta$ for fixed phase sum $(\phi_1 + \phi_2)/2$. 
 %(θ - (ϕ₁ + ϕ₂)/2) and (Θ - (ϕ₁ - ϕ₂)/2). 
This establishes a direct orthogonal mapping between our experimental control parameters and the two distinct coherences we aim to  independently probe.
%Therefore, the phases of the measurement pulses constitute the calibrated probe used to measure the coherence, and the contrast of the resulting interference fringes provides a quantitative measure of these distinct coherences.
%This mapping is conceptually represented by the perpendicular $\theta$ and $\Theta$ axes in 
Figure \ref{fig:concept}\textbf{e} conceptually represents this mapping with perpendicular $\theta$ and $\Theta$ axes.

Beyond these independent projections, the full two-dimensional interference landscape captures the state's {next order qubit-parity coherence and interference}, {upto $\ket{g, 2n} \leftrightarrow \ket{e, 2n \pm  3}$}. 
As detailed in Appendix~\ref{app:offdiag}, our two-pulse sequence is sensitive to coherence terms $|n\rangle\langle m|$ between Fock states of the same parity up to a separation of $|n - m| \leq 3$.
This sequence extends the single pulse verification by probing this third-order Fock coherences.
This range of Fock separation is sufficient for our purpose, which is to verify the creation of a non-classical state by measuring its quantum signatures. 
For instance, the two-pulse sequence is capable of probing third-order coherence terms, such as $|g, n\rangle\langle e, n-3|$, which are a blind spot for the single-pulse method.

\textit{Internal oscillator coherence}, the phase relationship \textit{within} a single parity subspace, probed by scanning the phase difference, $(\phi_1 - \phi_2)/2$, tests interference \textit{among the different Fock state components} (e.g. between $|g,n\rangle$ and $|g,n+2\rangle$) within a single parity subspace,  establishing a  signature of parity definiteness with multi-phonon components  in the motional state. 
%This makes the phase difference a specific probe for the internal oscillator coherence $\Theta$. 
This makes the phase difference a probe for specific orders of internal oscillator coherence $\Theta$. 

Any experimental state is subject to decoherence, which makes long-range coherences profoundly more fragile than short-range ones.
To quantitatively distinguish our prepared state from an incoherent mixture as detailed in Appendix~\ref{Appendix:decoherence}, we consider a phenomenological model: if the coherence between adjacent number states $|n\rangle\langle n-1|$ is described by a factor $w \le 1$, the coherence for a separation of $k=|n-m|$ would scale as $w^k$, decaying exponentially. 
In the prepared states, the long range coherence $|n - m| >3$ would be present {if $1-w\ll 1$}. While the current two pulse verification scheme is blind to them, {it can be used to estimate $w$.}
Our two-pulse sequence is therefore a {sufficient} probe for the short range coherences ($k \le 3$), which may be used to infer the long range coherence factors that would survive if $w \approx 1$. 
{Furthermore,} as detailed in Appendix \ref{appendix:equivalnece}, the total contrast measured over the full $(\phi_1, \phi_2)$ space {roughly} corresponds to the weighted sum of both qubit-oscillator and internal oscillator coherence magnitudes.
\textbf{Interference quantification.} We quantify both interferences by the single-pulse and two-pulse measurements using two standard metrics.
%\begin{itemize}
    The contrast $C(\bm{\varphi})\equiv P_{\text{max}}(\bm{\varphi}) - P_{\text{min}}(\bm{\varphi})$ measures the  absolute amplitude of the population oscillation% and  {$\av{C}_{\bm{\varphi}}$ its average over many realizations of the {interaction phases} $\bm{\varphi}$}
, where  $P_{\text{max}}(\bm{\varphi})$ and $P_{\text{min}}(\bm{\varphi})$ are the maximal and minimal values of a measured probability distribution $P_g$ from the state and measurement over $\theta$ and $\Theta$, or equivalently over $\bm{\phi}=(\phi_1,\phi_2)$. 
    %It is a fragile metric, highly sensitive to decoherence and, crucially, to the unknown preparation phase which may result in an unbalanced interferometric pathways.
  %   This metric is fragile because it depends on the specific motional state prepared, which is sensitive to the unknown phase. 
  It is sensitive to both the underlying coherence and the overlap between the prepared state and the measurement basis. 
  Low contrast can result from either decoherence or a non-optimal measurement setting, even for a pure state.
     % An {suboptimal} phase can lead to unbalanced interferometric pathways and thus low contrast, even for a perfectly pure state.
    The normalized contrast, or visibility $V(\bm{\varphi})\equiv (P_{\text{max}}(\bm{\varphi}) - P_{\text{min}}(\bm{\varphi}))/(P_{\text{max}}(\bm{\varphi}) + P_{\text{min}}(\bm{\varphi}))$
    is fundamentally a measure of the clarity of the interference fringe.  
    \cmmnt{Crucially, high visibility can persist even in the presence of decoherence, provided the interfering pathways remain balanced, whereas contrast decays. }
    %Therefore, the visibility serves as our primary experimental quantifier of the underlying quantum coherence,    whose limits are dictated by the spectral mismatch between the state and measurement basis.
{Crucially, high visibility normalizes against \textit{global population loss} provided the interfering pathways remain balanced; however, it remains strictly bounded by the state's \textit{internal coherence}, as stochastic dephasing eventually fills the interference minima and suppresses visibility. 
Therefore, visibility serves as our primary quantifier of coherence. Its upper bound is constrained  by the \textit{spectral mismatch}—the imperfect overlap between the prepared state's Fock-state distribution and the states projected by the verification pulses \cite{Leibfried2003, TurchettePRA2000Decoherence}. This mismatch in number-state support limits the maximum achievable visibility even for ideal pure states. }
  To illustrate, consider an interference experiment where a prepared state with motional components $\ket{\psi_\mathrm{odd}(\bm{\varphi})}$ and $\ket{\psi_\mathrm{even}(\bm{\varphi})}$ in (\ref{eq:psi}) is projected onto a measurement basis with components $\ket{\psi^\text{(measure)}_\mathrm{odd}(\bm{\phi})}$ and $\ket{\psi^\text{(measure)}_\mathrm{even}(\bm{\phi})}$ in (\ref{eq:psim}). 
  The contrast and visibility, for the prepared balanced states, are determined by the overlap between the prepared state $|\Psi\rangle_{\theta,\Theta}$ in Eq. (\ref{eq:psiTheta}) and the measurement basis $|\Psi_m\rangle$  in Eq.~(\ref{eq:psim}). %The ground state population modulates as $P_g = \frac{1}{4} | \langle \psi_{\text{even}}^{\text{(measure)}} | \psi_{\text{even}}(\Theta) \rangle + e^{i\theta} \langle \psi_{\text{odd}}^{\text{(measure)}} | \psi_{\text{odd}}(\Theta) \rangle |^2$. 
 Defining the overlaps of even and odd components  as $\alpha_e(\bm{\varphi}) = \langle \psi_{\text{even}}^{\text{(measure)}}(\phi_1, \phi_2) | \psi_{\text{even}}(\bm{\varphi}, \Theta) \rangle$ and $\alpha_o(\bm{\varphi}) = \langle \psi_{\text{odd}}^{\text{(measure)}}(\phi_1, \phi_2) | \psi_{\text{odd}}(\bm{\varphi}, \Theta) \rangle$, the contrast and visibility about $\theta$ are $C(\bm{\varphi}) = |\alpha_e(\bm{\varphi}) \alpha_o^*(\bm{\varphi})|$ and $V(\bm{\varphi}) = 2|\alpha_e(\bm{\varphi}) \alpha_o^*(\bm{\varphi})| / (|\alpha_e(\bm{\varphi})|^2 + |\alpha_o(\bm{\varphi})|^2)$. 
  The visibility is maximized when the magnitudes of the interfering pathway overlaps are {balanced}: $\left|\braketk{\psi^\text{(measure)}_\mathrm{even}(\bm{\phi})}{\psi_\mathrm{odd}(\bm{\varphi})}\right| = \left|\braketk{\psi^\text{(measure)}_\mathrm{odd}(\bm{\phi})}{\psi_\mathrm{even}(\bm{\varphi})}\right|$. 
    {High visibility (approaching $1$), which indicates such balanced interfering pathways,}  allows for a perfectly dark output {(zero population) at the interference minimum} in a qubit-oscillator {interference}. 
  %This is because visibility is insensitive to the phase-induced amplitude variations that affect the contrast. 
  %High visibility (approaching $1$) {also} indicates that the interfering pathways are perfectly balanced.
   % {Crucially, it is insensitive to the phase-induced amplitude variations that affect the contrast.}
 In order to account for \cmmnt{stable but unknown} interaction parameters, we average the contrast and visibility over many realizations of the interaction phases $\bm{\varphi}$ {during the state preparation}, throughout the work as 
 \begin{align}
     \av{C}=\frac{1}{(2\pi)^N}\int_{\bm{\varphi}} d\bm{\varphi} ~ C(\bm{\varphi}), ~~
     \av{V}=\frac{1}{(2\pi)^N}\int_{\bm{\varphi}} d\bm{\varphi} ~ V(\bm{\varphi}) \label{eq:avCV},
 \end{align}
 where the integration is over all possible parameters in $\bm{\varphi}$.  We computed the integral numerically using a Monte Carlo method, for which the averages saturate rapidly with the number of samples, and experimentally over larger than $100$ realizations.

\section*{Results}
%{We first characterize the generated motional states by their phonon statistics and fundamental structure, and then quantify their coherence using single- and two-pulse interferometry.}

\subparagraph{Measurement of Oscillator Statistics}

\begin{figure}[ht!]   \centering   % USER NOTE: This combines your old Figure 2a and 2b.   % The labels and references need to point here now.   
\includegraphics[width=\textwidth]{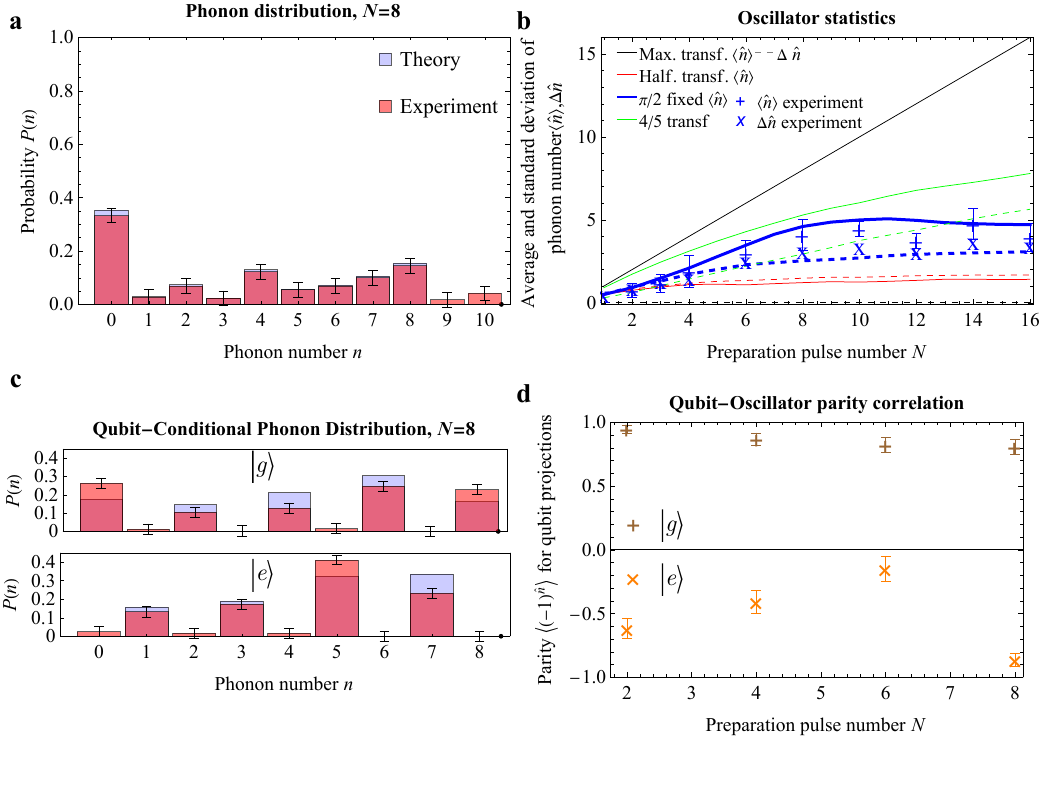} % Placeholder for your combined 2a and 2b plots  
\caption{\textbf{Oscillator statistics.} 
\textbf{a} Phonon number distribution $P(n)$ for a state prepared with $N=8$ pulses. The plot compares the experimentally measured distribution (red) against the theoretical prediction (light blue, see Appendix \ref{app:reconstruction}), showing the creation of a {dispersed} motional state.
\textbf{b} Growth of the mean phonon number $\langle \hat{n} \rangle$ and its standard deviation $\Delta \hat{n}$ with the number of preparation pulses $N$. This confirms the generation of increasingly populated and {adequately} delocalized motional states. 
\textbf{c} Conditional phonon number distributions demonstrating the strong qubit-parity correlation {in the dispersed $P(n)$}. 
%The differences between the experiment and theory are shown in orange {and light blue} based on its sign. 
{Deviations between experiment and theory are color-coded by sign (orange: positive, light blue: negative).}
(Top) the distribution when the qubit is measured in the ground state $|g\rangle$, which is composed of {dominantly} even phonon numbers. (Bottom) The distribution for the excited state $|e\rangle$, composed of {dominantly} odd phonon numbers.
\textbf{d} Measurement of the qubit-parity correlation. 
The average motional parity $\langle(-1)^{\hat{n}}\rangle$ as a function of the preparation pulse number, $N$. 
      %  The measurement is conditioned on projectively measuring the qubit in the ground state $|g\rangle$  or the excited state $|e\rangle$. 
The \cmmnt{selected} data confirms that the ground (excited) qubit state is correlated with even (odd) phonon numbers, demonstrating the  imprinting of the desired state structure. 
%{(The $N=6$ data point for the excited state parity is omitted because low population in $|e\rangle$ prevented statistically significant parity estimation to overcome the projection noise during post-selection. See Appendix \ref{appendix:experiment}).}
% \textbf{b} {Growth of state complexity, quantified by the Shannon entropy $S$ of the phonon distribution, as a function of the preparation pulse number $N$. The clear increase in entropy for the experimental fixed $\pi/2$ pulse method (blue) confirms the state's population is distributed across a larger number of Fock states. The results are compared with theoretical predictions for an ideal half-transfer (red) and a 4/5 transfer (green) scenario.}
 }   \label{fig:new_fig2}
\end{figure}

{Rabi flopping (Fig. \ref{fig:concept}\textbf{c}) reveals motional excitation increasing with $N$, saturating at $\langle \hat{n} \rangle \approx \Delta \hat{n} \approx 4$. The ratio $\langle \hat{n} \rangle/\Delta \hat{n} \approx 1$ arises from the fixed-duration pulse choice (Fig. \ref{fig:new_fig2}\textbf{a}), validating the deterministic generation of complex motional states.}
\cmmnt{Analysis of the phonon distribution (Fig. \ref{fig:concept}\textbf{c}), obtained from Rabi flopping measurements, confirms {that we} generate states with increasing motional excitation, saturating at an average phonon number $\langle \hat{n} \rangle \ge 4$ and a significant standard deviation $\Delta \hat{n}\approx \langle \hat{n} \rangle$.
We confirm that our {experiment} generates increasingly complex motional states with the intended structure. By applying a varied number of preparation pulses $N$, we measure the resulting phonon distribution $P(n)$ {for Fock number $n$} via Rabi flopping (see Fig. \ref{fig:concept}c). 
The mean phonon number $\langle \hat{n} \rangle$ and its standard deviation $\Delta \hat{n}$ grows with $N$, saturating around $\langle \hat{n} \rangle \approx 4$,  and thus $\langle \hat{n} \rangle/\Delta \hat{n}\approx 1$ for our choice of fixed-duration pulses (see  Fig. \ref{fig:new_fig2}\textbf{a}).} 
%This confirms the creation of large, delocalized motional states, which saturate at a mean phonon number $\langle n \rangle \approx 4$.

%-- START: NEW FIGURE 2 (Formerly Figure 2a, 2b) ---

%The generation of states with significant motional excitation and {its spread} is confirmed by analyzing the phonon distribution $P(n)$. 
Figure~\ref{fig:new_fig2}\textbf{a} shows the average phonon number $\langle \hat{n} \rangle$, its standard deviation $\Delta \hat{n}$ {and their saturation to a value of approximately $4$ at large $N$, a consequence of the fixed pulse length. }
%The saturation of $\langle \hat{n} \rangle$ and $\Delta \hat{n}$ at large $N$ (Fig.~\ref{fig:new_fig2}\textbf{a}) {to a value of approximately $4$} is a consequence of the fixed pulse length, not a fundamental limitation of the {experiment}. 
As shown by the simulation in Fig. \ref{fig:new_fig2}\textbf{a}, dynamically adjusting the pulse durations to account for the $\sqrt{n}$-dependence of the Rabi frequency would allow for continued growth of the motional state size. 
 Since the sideband Rabi frequency depends on the phonon number $n$, our fixed-duration pulse eventually acts as a $2\pi$ pulse for a specific high-$n$ transition. 
 This completes a full Rabi cycle, preventing further population transfer and capping the motional excitation. 
 %This saturation point can be shifted to higher $n$ simply by choosing a different pulse length.
 %The {observed saturation at $\langle \hat{n} \rangle \approx 4$ is a predictable consequence of fixed-area pulses reaching $2\pi$ cycles for specific $n$ transitions. 
 This saturation is an experimental byproduct of {fixed-area pulses} rather than a fundamental limitation of the protocol's scalability, and a larger motional states can be generated simply by adjusting the pulse duration.
 
% To provide a more complete picture of the state's delocalization in the Fock basis, we also calculate the Shannon entropy of the $P(n)$ distribution, defined as $S = -\sum_{n} P(n) \log_2 P(n)$ (Fig.~\ref{fig:new_fig2}\textbf{b}). 
% The clear increase in entropy with the number of pulses $N$, {saturating for $N \gtrsim 10$,} provides direct evidence that the state's population is being distributed across a progressively larger number of Fock states, confirming an increase in the state's complexity beyond what is captured by the variance alone.

% \subparagraph{Qubit-Parity Correlation}

% \begin{figure}
%     \centering
%     \includegraphics[width=0.5\linewidth]{parity14Oct2025.pdf}
%     \caption{ {\textbf{Measurement of the qubit-parity correlation.} 
%         The average motional parity $\langle(-1)^{\hat{n}}\rangle$ as a function of the preparation pulse number, $N$. 
%         The measurement is conditioned on projectively measuring the qubit in the ground state $|g\rangle$  or the excited state $|e\rangle$. 
%         The data confirms that the ground (excited) qubit state is strongly correlated with even (odd) phonon numbers, demonstrating the successful imprinting of the desired state structure. }
%     }
%     \label{fig:parity}
% \end{figure}

A core feature of our {experiment} is the deterministic generation of a state with a {strong} qubit-parity correlation. 
%To verify this foundational principle, we projectively measure the qubit state and then analyze the parity of the conditional motional state. 
Crucially, we verify the foundational qubit-parity correlation. 
As shown {conceptually} in Fig. \ref{fig:concept}\textbf{d}, projective measurements {together with Rabi flopping} confirm the expected parity-qubit association across all preparation steps.
While ideal theory predicts a perfect, strict correlation, experimental factors  may lead to a small degradation of this structure, as seen in Fig \ref{fig:new_fig2}c.
The results, presented in Figure~\ref{fig:new_fig2}\textbf{d}, confirm the preservation of the  parity-qubit lock. The decoherence model detailed in Appendix \ref{Appendix:decoherence} indicates that the system maintains a high degree of purity. %provide strong evidence of this correlation. 
We confirm that conditioning on the ground state $\ket{g}$ consistently yields a positive average parity $\langle(-1)^{\hat{n}}\rangle$, indicating that the motional state is composed of {dominantly} even phonon numbers. 
Conversely, conditioning on the excited state $\ket{e}$ yields a negative average parity, {although its correlation with odd phonon numbers were only weakly manifested for data sets with $N\le6$}. 
{We note that these reported odd-parity values  are affected by increased photodetection noise, as the odd components are less populated in the prepared superposition states; this effect becomes more pronounced for smaller $N$. In addition, the population {measurements} are highly sensitive to the durations of the probing Rabi pulses, which further {reduces} the evaluated parity values. Consequently, the reported results should be regarded as a lower bound on the {average parity} of the prepared states.}
This imprinting of the desired structure demonstrates that the alternating sideband pulse sequence works as intended, enforcing the qubit-parity correlation that is robust to the unknown-but-stable interaction phases.

\subparagraph{Interference by single- and two-pulse measurements}

\begin{figure}
    \centering
    \includegraphics[width=1.0\linewidth]{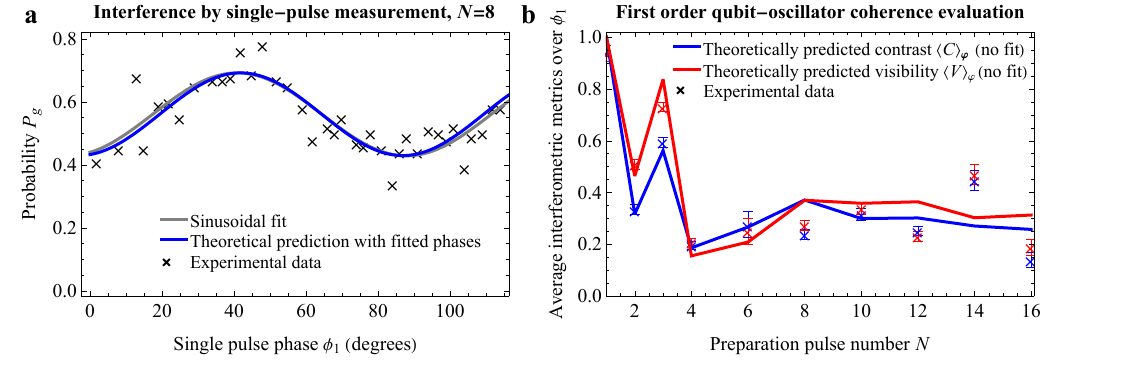}
    \caption{\textbf{Interference by single-pulse measurement.} \textbf{a} Interference fringe from a single-pulse verification measurement for $N=8$. The final ground state population, $P_g$, is plotted as a function of the verification pulse phase, $\phi_1$. 
    {The theoretical curve represents a fit to this specific dataset, using the unknown preparation phases as parameters to validate the accuracy of our model for a single realization.}
    %The clear sinusoidal modulation (solid line) fitting the experimental data (crosses) confirms the presence of quantum coherence in the prepared state. 
    \textbf{b} Contrast and visibility of the final qubit population modulation $P_g$ from a single-pulse verification scan. 
    {The theoretical prediction curves, which require no fitting, are averaged over all possible unknown stable phases to show the expected performance.}
    %This shows strong interference in an  agreement with theory predictions without fitted phases. 
    This non-zero contrast confirms the presence of coherence in the prepared state, distinguishing it from a simple incoherent mixture {(\ref{eq:incoherentDM2}) or} (\ref{eq:incoherentDM}). %However, this probe cannot by itself distinguish the type of coherence.
    }
    \label{fig:coherence}
\end{figure}

To validate our {experiment}, we employ a two-way approach for theoretical comparisons. 
First, to confirm the accuracy of our underlying physical model, we show that it can precisely reproduce the data from a specific experimental run by treating the unknown-but-stable interaction phases ($\phi_i$) as fitting parameters. 
Second, to demonstrate the general robustness of our {experiment}, we compare the experimental metrics against theoretical predictions that have been averaged over all possible {(practically randomly sampled)} interaction phases. 
This second method requires no fitting and directly tests the expected average-case performance.

Figure~\ref{fig:coherence} shows the measured contrast of this modulation versus the number of preparation pulses $N$ obtained from a single-pulse verification scan. The strong interference confirms that coherence is successfully generated and maintained, ruling out a fully mixed state. 
%Figure \ref{fig:macroscopicitynbarvsN}\textbf{b}  plots the contrast of the qubit population modulation, obtained from a single-pulse verification scan, as a function of the number of preparation pulses $N$. 
This contrast directly measures the interference between the qubit and the oscillator's parity subspaces, reflecting their mutual coherence. 
The plot compares our experimental data (blue crosses) with theoretical predictions for two different pulse strategies: our fixed $\pi/2$ pulse method (blue squares) and an ideal half-transfer scenario (red triangles). 
The experimental results follow the general trend of the fixed $\pi/2$ model, showing a decrease in contrast as the state's complexity (and mean phonon number) grows{, before leveling off at a saturated value of $\sim0.2$ for $N \ge 8$}. 
The observed saturation and the deviation from the ideal half-transfer case, particularly at larger $N$, can be attributed to a combination of two factors: the specific state and detection properties by our pulse sequence, and the accumulation of decoherence from environmental noise sources during the longer pulse sequences. 
This general decoherence, distinct from the stable phase error our {experiment} is designed {specifically} to resist, reduces the overall coherent fraction of the state, thereby lowering the absolute contrast.

\begin{figure}
    \centering
    \includegraphics[width=1.0\textwidth]{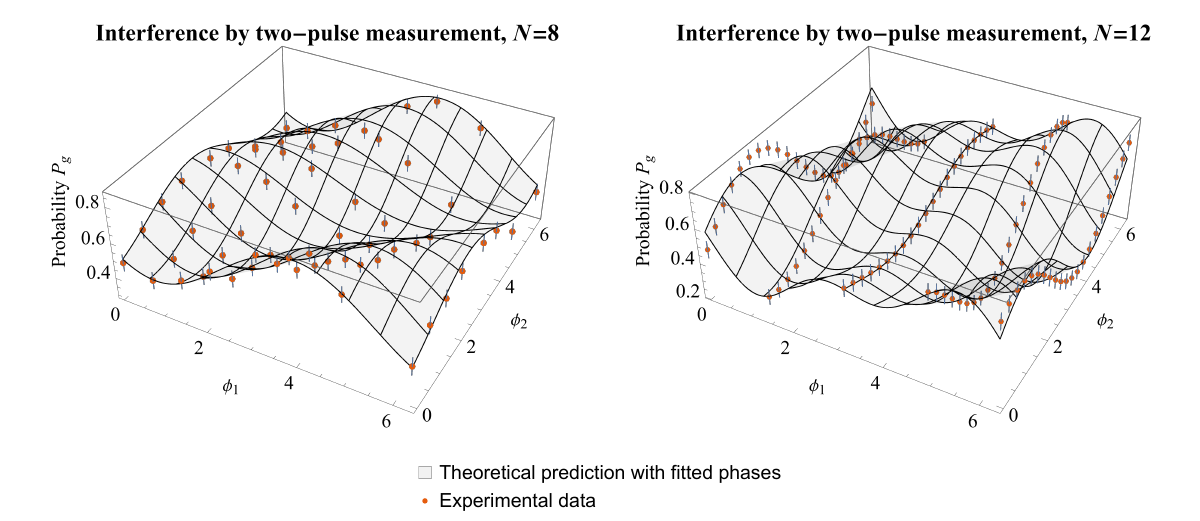}
    \caption{
    \textbf{Two-dimensional interference landscape} for \textbf{a} $N=8$ and \textbf{b} $N=12$ pulses. 
    Experimental data (points) are overlaid with a theoretical model {where the interaction phases have been fitted to these specific runs. The excellent agreement validates the model's ability to describe individual experimental outcomes.} %using fitted interaction phases.
    % \textbf{a}~Experimentally measured two-dimensional interference pattern, with a theoretical prediction {with fitted phases for $N=8$ with a theoretical prediction from an ideal model (excluding decoherence) using fitted interaction phases}. The ground state probability $P_g$ is plotted as a function of the two verification pulse phases, $\phi_1$ and $\phi_2$. 
    % {\textbf{b} An equivalent interference for $N=12$.}
         %Experimental data points (red markers with error bars) are overlaid on the theoretical surface plot, demonstrating a strong agreement and confirming the complex coherence structure of the prepared state.
    %Measured excited state probability $P_g$ as a function of the two verification pulse phases, $\phi_1$ and $\phi_2$, for $N=8$. The clear modulation demonstrates the state's coherence, matching the theoretical surface plot.
    }
    \label{fig:Pevsphi}
\end{figure}

\begin{figure}
    \centering
    \includegraphics[width=1.0\linewidth]{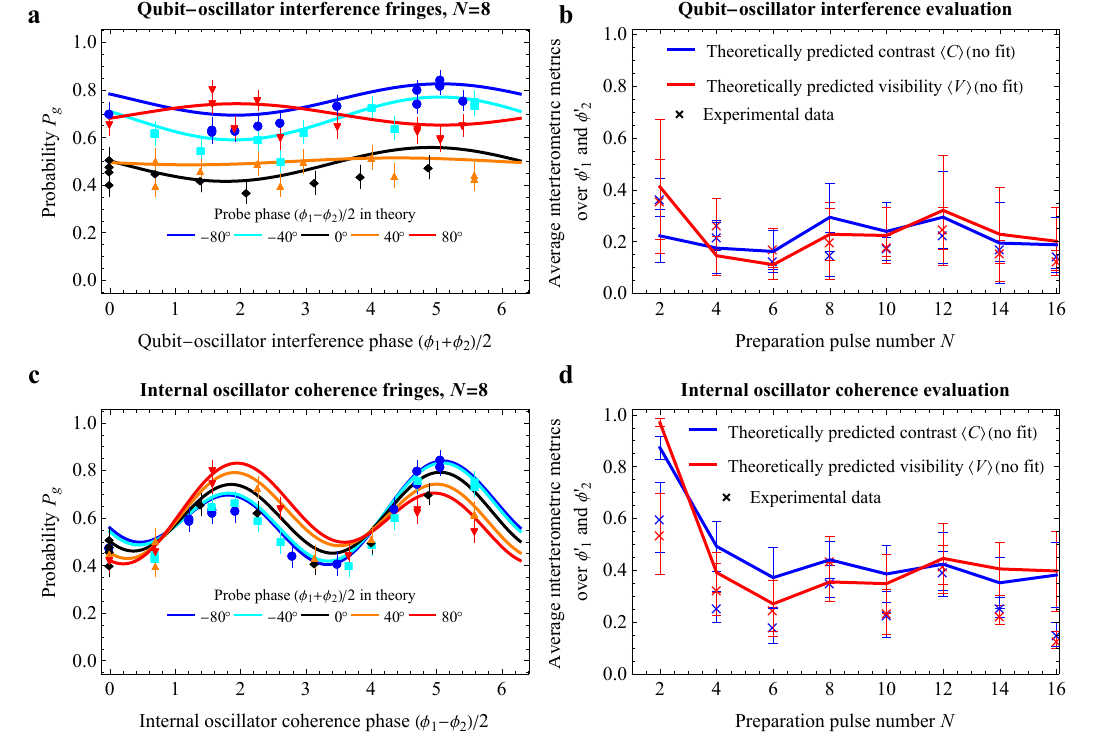}
   % \includegraphics[width=0.496\linewidth]{fig2a 24 Feb 2025.pdf}
   % \includegraphics[width=0.496 \linewidth]{fig2b 24 Feb 2025.pdf}
   % \includegraphics[width=0.5\linewidth]{fig2d 24 Feb 2025.pdf}
%     \caption{
% Macroscopicity growth through preparation pulses. We consider three scenarios: (1) generation of Fock state $\ket{N}$ (achieved by full population transfer), (2) transferring half of the population in the largest Fock space one step higher at each stage, and (3) using a fixed pulse length $t=\frac{\pi}{2}$. The final scenario, which is our primary focus, shows significant growth in the average phonon number $\langle \hat{n} \rangle$ and its standard deviation. If the strength is randomly chosen, it grows similarly to the fixed pulse strength case.
% }
\caption{\textbf{Interference by two-pulse measurements.} 
(\textbf{a, c}) Interference fringes for $N=8$, showing  agreement between experimental data and our theoretical model when the unknown interaction phases $\bm{\varphi}$ are fitted to the specific dataset.
(\textbf{b, d}) Average interferometric metrics versus preparation pulse number $N$. Experimental data is compared against {parameter-free} {ideal} theoretical predictions that have been averaged over randomly sampled all possible interaction phases. The close agreement with the theoretical visibility fundamentally bounded (to the theoretical value $\lesssim 0.5$)  validates the general robustness to unknown phases. This effect is present in both the experimental data and the theoretical predictions.
    }
    \label{fig:macroscopicitynbarvsN}
\end{figure}

Figure~\ref{fig:Pevsphi} provides a detailed visualization of the two-dimensional interference landscape for a state prepared with $N=8$ pulses, which serves as the foundational data for the summary metrics presented in Figure~\ref{fig:macroscopicitynbarvsN}. Specifically, Figure~\ref{fig:Pevsphi} shows the experimentally measured ground state probability, $P_g$, as a function of the two independent verification phases, $\phi_1$ and $\phi_2$. This two-dimensional data is then projected onto two orthogonal axes to independently analyze the two distinct types of coherence. The projection onto the phase sum axis, $ (\phi_1 + \phi_2)/2$, isolates the interference fringes that are a direct signature of the \textit{qubit-oscillator coherence} (Figure~\ref{fig:macroscopicitynbarvsN}\textbf{a}). Conversely, the projection onto the phase difference axis, $ (\phi_1 - \phi_2)/2$, isolates the interference fringes that are a direct signature of the \textit{internal oscillator coherence}, depicted in Figure~\ref{fig:macroscopicitynbarvsN}\textbf{c}. From each of these one-dimensional interference fringes, the maximum ($P_{\max}$) and minimum ($P_{\min}$) population values are extracted. These values are then used to compute the two key interferometric metrics, the {averaged} contrast and the visibility (\ref{eq:avCV}). %Figure~\ref{fig:macroscopicitynbarvsN} is  generated by repeating this entire analysis procedure for states prepared with a varying number of pulses, $N$. Each data point in Figure~\ref{fig:macroscopicitynbarvsN} thus represents the average contrast and visibility calculated from a full two-dimensional scan. %, with Figure~\ref{fig:Pevsphi} illustrating the complete analysis for the specific case of $N=8$.

%To definitively characterize the prepared state, we employ a two-pulse verification scan that independently probes the two fundamental types of coherence. 
The results are presented in Figure~\ref{fig:macroscopicitynbarvsN}, which plots the average interference contrast $\av{C}$  and visibility $\av{V}$ (\ref{eq:avCV}) against $N$.
\cmmnt{These  metrics serve as our quantitative measure of the distinct types of coherence.}
Figure~\ref{fig:macroscopicitynbarvsN}\textbf{b} isolates the qubit-oscillator interference (probed by the phase half sum $\frac{\phi_1 + \phi_2}{2}$), while Figure~\ref{fig:macroscopicitynbarvsN}\textbf{d} probes the internal oscillator coherence (probed by the phase half difference $\frac{\phi_1 - \phi_2}{2}$). In both independent measurements, the experimental data for contrast and visibility closely track their respective theoretical predictions {without fitting}. A crucial aspect of this analysis is the theoretical maximum visibility {matching the theoretical limits of} $\approx 0.4$ for internal oscillator interference, while $\approx 0.2$ for the qubit-oscillator interference,  a ceiling imposed by the inherent mismatch between our prepared state and the measurement basis. 
%Achieving a visibility near this ideal limit is therefore a strong confirmation of the protocol's high fidelity.
%{For the internal oscillator coherence, the dominant contribution of the {averaged} contrast and visibility comes from the second order coherence $\ket{n}\bra{n-2}$ for all $n$'s.}
For the internal oscillator coherence, the dominant contribution to the 
averaged contrast and visibility comes from the second-order coherence 
terms $|\langle n | \rho_O | n-2 \rangle|$, where $\rho_O$ is the reduced 
density matrix of the oscillator.
The excellent agreement across both projections provides unambiguous evidence that the {experiment} generates and maintains a complex quantum state with both types of coherence. 
%This confirms that the observed decay at higher $N$ is not a failure of the protocol's core principle but a quantifiable effect of accumulating experimental imperfections, which our model correctly predicts.
This decay in coherence also saturates for $N \gtrsim 8$,  {matching the theoretical prediction}, a behavior interlinked with the saturation of the state's complexity seen in Fig.~\ref{fig:new_fig2}: once the state stops growing, the rate of accumulating decoherence also stabilizes, leading to the observed plateau in the coherence metrics.
The observed decay at higher $N$ deviating from the theory {can be attributed to}  a quantifiable effect of accumulating experimental imperfections. 
The increased sequence duration for larger $N$ allows more time for time-dependent decoherence, such as motional heating from electric-field noise and qubit dephasing from magnetic-field fluctuations. 
Furthermore, each of the $N$ control pulses contributes a small random error from laser fluctuations, and these errors compound, reducing the final state fidelity.
A key feature of the data in Fig.~\ref{fig:macroscopicitynbarvsN} is the near-identical measured average contrast $\langle C \rangle$ and visibility $\langle V \rangle$, especially for a low number of preparation pulses $N$. From their definitions, the condition {$\av{C} \approx \av{V}$} mathematically implies that $P_{\text{max}} + P_{\text{min}} \approx 1$ {for various realizations}. This is a direct signature of a well-balanced interferometric measurement, where the two interfering quantum pathways---defined by the projection of the prepared state onto the measurement basis---have nearly equal effective amplitudes. 
%This demonstrates that our verification protocol is well-matched to the prepared states, allowing for a clear observation of interference. 
%The slight divergence between contrast and visibility suggests the onset of imperfections that break this symmetry, such as the accumulation of a small incoherent background population.
Notably, the experimental visibility remains close to the contrast for both {single- and two-pulse} measurements {as in Figs \ref{fig:coherence} and \ref{fig:macroscopicitynbarvsN}}. %, signifying that the interference fringes are of high quality with a low incoherent background, and also that the measurement is not optimized for the visibility that requires a significant asymmetric pathways. 

A key feature in Fig.~\ref{fig:macroscopicitynbarvsN} is the higher measured visibility for internal oscillator coherence compared to qubit-oscillator coherence. %This asymmetry is not a signature of superior intrinsic coherence, but rather an inherent feature of our measurement protocol. 
The theoretical model for a general qubit-oscillator correlated states not bound to any preparation, including qubit-parity correlated states, imposes higher maximum visibilities and contrasts for internal oscillator coherence than those for qubit-oscillator coherence. 
%Our experiment's close agreement with these distinct ceilings validates the model and confirms the protocol's high fidelity.
{To quantify the state purity beyond raw visibility, we find the purity about $ \approx 0.82$ {($w\gtrsim 0.9$)} as in Appendix \ref{Appendix:decoherence}, indicating that the observed contrast is nearly optimal and the system maintains a high degree of underlying quantum coherence despite the spectral mismatch with the measurement basis.}

\section*{CONCLUSION}

We have proposed and experimentally demonstrated  generation and verification of qubit-{parity correlation and} interference in the presence of stable but unknown phases. 
% By enforcing a strict qubit-parity correlation with a sequence of alternating sideband pulses, we create states that are structurally {resilient against} unknown  stable phases of the preparation pulses. 
% Our verification method, which probes for phase-dependent oscillations in the qubit population, provides a structured and quantitative  measure of the resulting {total} coherence, comprised of qubit-oscillator coherence and {internal oscillator} coherence. 
% {We have experimentally demonstrated a protocol for generating complex qubit-oscillator entangled states that are structurally resilient to unknown stable interaction phases. }
%{We have experimentally demonstrated a protocol for generating complex qubit-oscillator entangled states that are structurally resilient to unknown stable interaction phases.}
% The resulting states exhibit a substantial interference visibility of approximately $0.4$. 
% Crucially, this value is not a measure of imperfection, but rather a near-perfect match to the maximum visibility predicted by ideal theoretical models {without fitting parameters}.  
The resulting states exhibit a substantial interference visibility and contrast of approximately $0.4$ {(theoretically $\approx 0.47$)} for internal oscillator coherence, and $0.2$ {(theoretically $\approx 0.22$)} for qubit-oscillator interference. Crucially, these values show strong agreement with the theoretical maximum visibility imposed by our measurement method.
The close correspondence demonstrates that the primary constraint on visibility is the overlap between our prepared state and measurement basis, rather than infidelity.
% This excellent agreement between experiment and theory is the signature of the protocol's success, confirming that genuine, complex quantum coherence is generated and preserved. 
% The primary limitations observed in this experiment arise not from a failure of this principle, but from the accumulation of general decoherence (e.g., from motional heating and field fluctuations) during longer pulse sequences. 
% This manifests as a small, quantifiable deviation from the ideal model in the interference contrast. 
% Our work thus demonstrates a robust design principle and provides a platform for studying decoherence in increasingly macroscopic quantum states.
%Our results demonstrate a strategic shift from active phase stabilization to {passive resilience by design.} 
%The measured visibility of ~0.4 is not a limitation but rather a quantitative measure of the coherence achieved under realistic conditions, including decoherence and pulse imperfections. 
%The measured visibility and contrast of $\approx 0.4$ is {not a sign of imperfection but is rather the key signature of our protocol's success— } {a direct, quantitative measure of the {high-fidelity} coherence that survives all realistic imperfections. 
This ceiling of visibility and contrast can be further enhanced by optimizing the verification pulse area.
The two-pulse {measurement} sequence provides a robust, unambiguous witness of {interference from hybrid coherence} that avoids potential false-positive visibilities from mixed states by a longer pulse scheme.
%This high visibility and contrast confirms the integrity of the underlying quantum state, and is the most crucial signature of our protocol's robustness.
The coherence decay during longer sequences arises from the saturation of the state's complexity, while the deviation from the theory occurs from general decoherence (e.g., from motional heating and magnetic field fluctuations) and the accumulation of small pulse-to-pulse control errors. 
{This experiment demonstrates {\textit{phase-insensitive robustness}} to stable-but-unknown phases that enables hardware-efficient qubit-parity quantum processing.}
Rather than relying on active correction, it can be exported to other platforms and more complex experiment which may inspire new approaches to building robust quantum components. %promising to accelerate the development of scalable {and robust} quantum technologies by significantly reducing control overhead. 
%Our work thus opens several immediate avenues for future research. 
%The most direct and practical next step is to apply this protocol to a collective motional mode of a multi-ion crystal. This might enable the robust generation of multi-qubit entangled states, such as GHZ states, whose creation is typically highly sensitive to the precise control of laser phases. 
Our work thus opens several immediate avenues for future research. It joins a broad effort into generating macroscopic quantum states, analogous to Schrödinger's cat, pioneered in cavity QED with microwave photons \cite{BruneHaroche1996PhysRevLett.77.4887Cat} and extended to numerous platforms, including propagating optical fields \cite{Hacker2019NatPhoDeterministic,Morin2014Remote,Jeong2014Generation}, integrated photonic architectures \cite{KnollmannOptica2024Integrated}, {or mechanics of trapped ions~\cite{HempelNatPho2013, Kienzler2016}}. 
A direct extension of our {experiment} would be to create ``multimode''  states by entangling the qubit with several distinct motional modes of the same ion, a key step towards engineering complex oscillator states \cite{McCormick2019, Kim:PhononicNetwork}. 
%Furthermore, our focus on robustness to unknown \textit{control phases} is complementary to recent work demonstrating coherence generation from an imperfect \textit{initial state} \cite{Yang2025SciAdv2025HotCat}. 
{While recent work has demonstrated resilience to initial state imperfections (thermal noise) \cite{Yang2025SciAdv2025HotCat}, our approach addresses a complementary challenge: resilience to uncontrolled interaction dynamics. 
Together, these methods provide a pathway for robust quantum control in non-ideal environments.}
%Combining these approaches to achieve resilience against both control and state imperfections presents a compelling research direction. 
Building on these foundations, the principle of phase-insensitive entanglement could then be applied to a collective motional mode of a multi-ion crystal. This would provide a powerful tool for the robust generation of multi-qubit entangled states, 
such as GHZ states \cite{CiracZoller1995, Monz2011PhysRevLett.106.130506entanglement}, whose creation is typically highly sensitive to the precise control of laser phases.
Furthermore, this technique provides a powerful tool for creating and manipulating unexplored Schrödinger's cat states, for example by using higher order sideband interactions.
%Future work could focus on optimizing pulse sequences to maximize the achievable coherence, quantifying the ultimate limits imposed by decoherence, and adapting this robust design principle to other quantum platforms, such as superconducting circuits. 
This would broaden the accessibility of complex quantum phenomena for both fundamental exploration and, potentially, technological application.

\section*{Acknowledgment}

%K.P. has been supported by 21-13265X of the Czech Science Fundation. 
K.S., K.P., L.S. and R.F. have been supported by the project GA22-27431S of the Czech Science Foundation.
K.S., K.P, V.S., A.K., L.S. and R.F. also acknolwedge project CZ.02.01.01/00/22\_008/0004649 (QUEENTEC) of EU and MEYS Czech Republic.
R.F. also used support from the project No. LUC25006 of MEYS Czech Republic.

\section*{Author contributions}

{R.F. developed the theory idea and conceived the project. 
K.P., R.F., and L.S. developed the idea about the measurement and verification, and K.P. performed analytical and numerical calculations with inputs about evaluation and interpretation from R.F.. 
K.P. wrote the manuscript with inputs from R.F. and contributions of all authors.
K.S. and V.S. performed the experiment and analyzed the experimental data under the supervision of L.S..}

\FloatBarrier % <--- THIS IS THE WALL. No figures from above can pass this point.

\setcounter{figure}{0}
\numberwithin{figure}{section} 
\appendix
%\begin{appendices}

\section{Data Reconstruction with a decoherence Model}
\label{app:reconstruction}

While these panels show the coherence projected onto specific measurement axes, the state's total coherence is manifested by the maximum interference amplitude (the global maximum of $P_g$ minus the global minimum) over the full two-dimensional phase space of $(\phi_1, \phi_2)$, as in Appendix \ref{appendix:equivalnece}. 
As quantitatively validated in Appendix \ref{appendix:equivalnece} for general parity-entangled states, a Fourier decomposition of this signal $P_g(\phi_1, \phi_2)$ allows us to isolate and measure specific, multi-order coherence terms ($|n\rangle\langle m|$), and the contrast over verification pulse phases measures the total coherence, up to certain order $|m-n|\le 3$.

\begin{figure}
 \includegraphics[width=1.0\textwidth]{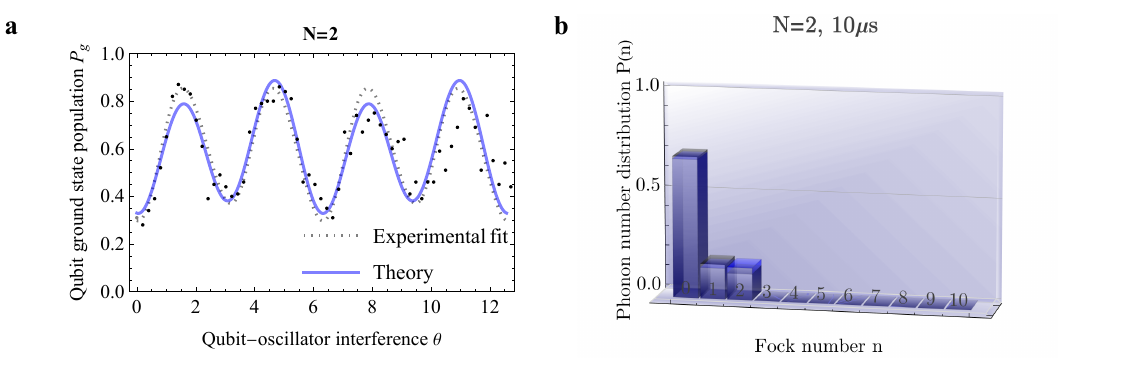}
 \caption{
{ \textbf{Reconstruction of experimental data for an $N=2$ state.}
The experimental data (points) are successfully reproduced by a theoretical model (solid lines and surface) incorporating a single fitted decoherence parameter ($w \approx 0.95$).
\textbf{a} Ramsey-like fringe decay, showing coherence.
\textbf{b} Reconstructed phonon number distribution $P(n)$ from Rabi flop measurements, with a mean phonon number $\langle \hat{n} \rangle = 0.475$.}
 %Reconstruction of experimental data. \textbf{a} Delay scan data. Reduction of off-diagonal element by decoherence was considered. 
 %\textbf{b} $P(n)$ data from Rabi flop. Mean phonon number is 0.475 here. 
 %\textbf{c} $P_e$ vs $\phi_1$ and $\phi_2$.
 %\textbf{c} The modulation of $P_g$ by varying phases of the two verification pulses, 1 red and 1 blue, as in the setup BR\textbar RB, \textbf{d} and in the setup BRBR\textbar RB. %Decoherence factor is found to be $w=0.242$.
 %\textbf{a,c,e} $P(n)$ data from Rabi flop and \textbf{b,d,f} contrast reproduction  (\textbf{a} $N=1$, \textbf{b} $N=2$, \textbf{a} $N=3$).
 }
 \label{fig:reconstruction}
\end{figure}

% \begin{figure}
%     \centering
%     \includegraphics[width=1.0\textwidth]{fig 4 3d plot.pdf}
%     \caption{Measured excited state probability $P_g$ as a function of the two verification pulse phases, $\phi_1$ and $\phi_2$, for $N=8$. The clear modulation demonstrates the state's coherence, matching the theoretical surface plot.}
%     \label{fig:Pevsphi}
% \end{figure}

\begin{figure}
 \includegraphics[width=1.0\textwidth]{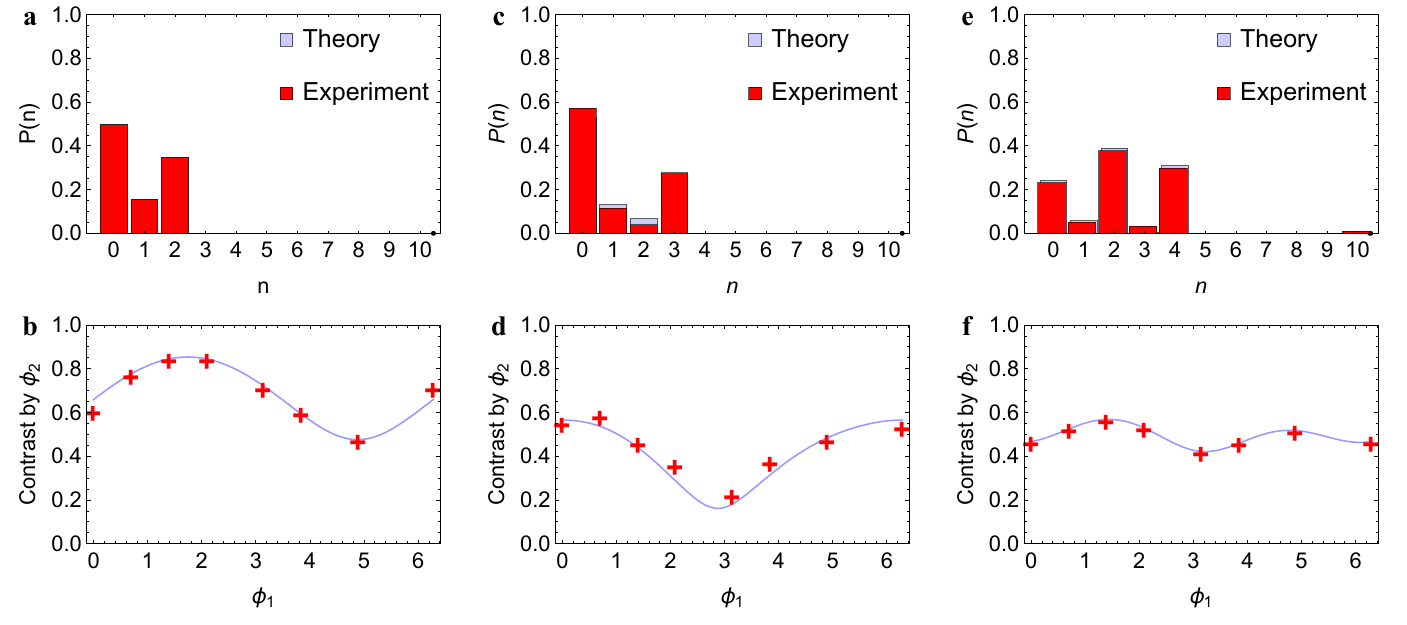}
 
 \caption{
 %Reconstruction of experimental data. \textbf{a} Delay scan data. Reduction of off-diagonal element by decoherence was considered. 
 %\textbf{b} $P(n)$ data from Rabi flop. Mean phonon number is 0.475 here. \textbf{c} The modulation of $P_g$ by varying phases of the two verification pulses, 1 red and 1 blue, as in the setup BR\textbar RB, \textbf{d} and in the setup BRBR\textbar RB. %Decoherence factor is found to be $w=0.242$.
 \textbf{a,c,e} $P(n)$ data from experiments and from theoretical description { (using fitted parameters)}, and \textbf{b,d,f} {corresponding $P_g$ modulation contrast vs $\phi_1$ (averaged over $\phi_2$) for various $N$} for various verification pulse phases.  (\textbf{a,b} $N=2$, \textbf{c,d} $N=3$, \textbf{e,f} $N=4$).
 }
 \label{fig:reconstruction2}
\end{figure}

We verify our  success by reconstructing the experimental data using a theoretical model that incorporates a single decoherence parameter, $w$ (see Eq. (\ref{eq:partialcoherent}) and Appendix \ref{Appendix:pulse}). 
As shown in Figures \ref{fig:reconstruction} and \ref{fig:Pevsphi}, this model faithfully reproduces the {experimental data}. 
%This excellent agreement between the data and the model, using a fitted coherence factor of $w \approx 0.9$, provides strong evidence that we prepared the intended entangled state, albeit with partial coherence.
Figure \ref{fig:reconstruction2} presents the measured phonon number distributions $P(n)$ and the corresponding visibility of the interference fringes for states prepared with $N=2, 3,$ and $4$ pulses. 
The visibility is plotted as a function of the first verification pulse phase, $\phi_1$, after averaging over the second phase, $\phi_2$. The solid lines represent the predictions from our theoretical model using a single fitted decoherence parameter, $w$, for each case. 
The excellent agreement between the experimental data (points) and the model (lines) across different state complexities validates our core claim: the experiment reliably generates the target qubit-parity entangled state, and its partial coherence can be accurately characterized.

\section{Minimal example of preparation and verification}

%\subparagraph{}
Our minimal experiment  outlines expandable steps demonstrating basic principles, ensuring  even basic experimental setups can clearly show coherence and increased macroscopicity.
{These principles extend to the longer sequences used in the experiment.}
Below, B (R) represents a blue (red) sideband pulse;  ``\textbar'' separates preparation and verification.

\textit{B\textbar R, B\textbar B ($N=1$):}
The minimal modulation demonstration in verification scans occurs even at $N=1$. 
Single blue pulses are used in both preparation and verification, with {the phase of the verification pulse being scanned}.   %Therefore, only the phase of the verification pulse is varied.
The input state $\ket{g,0}$ evolves after a preparation blue pulse (and global phase) to:
\begin{align}
\ket{\Psi}_\text{B}=\cos[\frac{t_1}{2}]\ket{g,0}+e^{\ii(\theta+\varphi_1)}\sin[\frac{t_1}{2}]\ket{e,1}.
\label{eq:PsiB}
\end{align}
$t_1$ is the preparation pulse length (area). 
The coherence phase $\theta$ is associated with the laser  phase $\phi_1$.

Scanning {the phase of a red verification pulse} does not create detection modulation, as it transfers population from  $\ket{e,1}$ to $\ket{g,2}$.
A red verification pulse (phase  \(\phi_1\) , length $t_R$) acting on $\ket{\Psi}_\text{B}$ yields:
\begin{align}
 &\ket{\Psi}_{B \vert R}=\cos \left[\frac{t_1}{2}\right]\ket{g,0}+e^{\ii (\theta+\phi_1)} \sin \left[\frac{t_1}{2}\right] \cos \left[\frac{t_R}{\sqrt{2}}\right]\ket{e,1}+\sin \left[\frac{t_1}{2}\right] e^{\ii \left(\theta+\varphi _1-\phi _1\right)} \sin \left[\frac{t_R}{\sqrt{2}}\right]\ket{g,2}.
\end{align}
%$P_g$ is not modulated by $\phi'_1$ due to the orthogonality of the oscillator states.
%After the verification blue pulse with phase $\phi_1'$, we obtain the state
Due to the orthogonality of the oscillator states,  qubit ground state weight $P_g$  is not modulated by $\phi_1$. 
In contrast,  a blue verification pulse (phase  $\phi_1$ and pulse length $t_B$) yields:
\begin{align}
 &\ket{\Psi}_{B \vert B}=\Big(e^{\ii (\theta+\phi_1)} \sin \left[\frac{t_1}{2}\right] \cos \left[\frac{t_B}{2}\right]-\cos \left[\frac{t_1}{2}\right] e^{\ii \phi_1} \sin \left[\frac{t_B}{2}\right]\Big)\ket{e,1}\nonumber\\
 &+\Big(\cos \left[\frac{t_1}{2}\right] \cos \left[\frac{t_B}{2}\right]+\sin \left[\frac{t_1}{2}\right] e^{\ii \left(\theta+\varphi_1-\phi_1\right)} \sin \left[\frac{t_B}{2}\right]\Big)\ket{g,0}.
\end{align}
% Due to the relative phase $\phi_1-\phi '_1$, the ground state probability $P_g$ modulates due to the interference between $\ket{g,0}$ and $\ket{e,1}$. 
% The maximum interference amplitude is found to be $1$ at $t_1=t'_1=\tfrac{\pi}{2}$, and the state in (\ref{eq:PsiB}) has the oscillator energy of $1/2$ in this case. 
% If incoherent state was prepared before the inverse pulse (that corresponds to the randomization of $\phi_1$), there will be no modulation.
Due to the relative phase $\varphi_1 - \phi_1$,  ground state probability $P_g$ modulates   due to interference between $\ket{g,0}$ and $\ket{e,1}$. Maximum interference amplitude is $1$ at $t_1 = t_{R,B} = \tfrac{\pi}{2}$, where the prepared state (\ref{eq:PsiB}) has an oscillator energy of $1/2$. If an incoherent state $\rho=\cos[\frac{t_1}{2}]^2\ket{g,0}\bra{g,0}+\sin[\frac{t_1}{2}]^2\ket{e,1}\bra{e,1}$ was prepared (instead of $\ket{\Psi}_\text{B}$), no modulation occurs.

% For partially coherent prepared state of the form $\rho_\Psi[w] \equiv w\ket{\Psi}\bra{\Psi} + (1-w)\operatorname{diag}[\ket{\Psi}\bra{\Psi}]$, where $\operatorname{diag}[\rho]$ is the matrix containing only the diagonal elements of $\rho$, the modulation amplitude is suppressed by a factor of $w$.
% The ground state probability is then given as
% \begin{align}
%     P_g=\frac{1}{2} \Big(1+\cos (t_1) \cos (t_1')-(1-w) \sin (t_1) \sin (t_1') \cos (\phi_1-\phi'_1)\Big).
% \end{align}
% Here, the modulation-independent term depend on preparation pulse length $t_j$ and verification pulse length $t_j'$, but not on $w$ or preparation pulse phases $\phi_j$. This is by definition the non-modulating term has to be equal to the contributions of the diagonal elements in the density matrix, which is not affected by the phase of the verification pulse. On the other hand,  the modulation amplitude $(1-w) \sin (t_1) \sin (t_1')$ depends linearly on $w$. This applies to all the examples below, and can be used to determine $w$ and $t_1'$. 
% For example, considering $t_j$ can be determined by the Fock weight distribution $P(n)$ prepared state that can be measured by Rabi flopping, we can precisely determine $t'_j$. From the modulation contrast, we can specify $w$.

  The {final} ground state probability $P_g$ is given by:
\begin{align}
    P_g = \frac{1}{2} \Big(1 + \cos(t_1) \cos(t_B) - w \sin(t_1) \sin(t_B) \cos(\varphi_1 - \phi_1)\Big).
\end{align}
% Here, the modulation-independent term depends on  preparation pulse length $t_j$ and  verification pulse length $t_j'$, but not on $w$ or the preparation pulse phases $\phi_j$   because, by definition, it corresponds to the contributions of the diagonal elements in the density matrix,  unaffected by the phase of the verification pulse. 
Here, the modulation amplitude is directly proportional to the coherence factor $w$. 
An ideal, perfectly coherent state ($w=1$) would yield the maximum possible visibility. 
The experimentally observed visibility of $~0.4$ therefore allows for a direct estimation of the state's coherence, indicating that $w \approx 0.9$. 
By independently measuring the state's populations, we can disentangle the loss of coherence (captured by $w$) from imperfections in the verification pulse itself.
% Conversely, the modulation amplitude $(1-w) \sin(t_1) \sin(t_1')$ depends linearly on $w$. 
% This principle applies to all the examples below and estimates $w$ and $t_1'$. 
% For example,  $t_j$ can be determined by  Fock state weight distribution $P(n)$, measured by Rabi flopping, we can precisely determine $t_j'$. From the modulation contrast with the knowledge of $t_j'$, we can then specify $w$.
{Conversely, the modulation amplitude depends linearly on the coherence factor $w$. 
This provides a general method for its determination: first, the Fock state populations $|c_n|^2 = P(n)$ are measured independently. 
From these populations and a given verification pulse length $t'$, the expected modulation amplitude for a perfectly coherent state ($w=1$) can be calculated, using Fourier series expansion. 
By comparing this ideal amplitude to the experimentally measured amplitude, we can directly estimate the true coherence factor $w$.}

%{
\textit{BR\textbar R, BR\textbar B ($N=2$) :} 
%After preparation pulses where $\phi_2$ is the phase of the second red pulse, the state generated is given as 
After preparation pulses with a second red pulse (phase $\phi_2$), the prepared state with the mutual phase encoding is:
\begin{align}
 \ket{\Psi}_\text{BR}=\cos \left[\frac{t_1}{2}\right]\ket{g,0}-\ii e^{\ii \left(\varphi _1+\theta+\frac{\pi }{2}\right)} \sin \left[\frac{t_1}{2}\right] \cos \left[\frac{t_2}{\sqrt{2}}\right]\ket{e,1}-e^{\ii \left(\varphi _1-\varphi _2\right)} \sin \left[\frac{t_1}{2}\right] \sin \left[\frac{t_2}{\sqrt{2}}\right]\ket{g,2}.
\label{eq:PsiBR}
\end{align}
After red verification pulse, the state evolves to
\begin{align}
  \ket{\Psi}_{BR\vert R}=&\cos \left[\frac{t_1}{2}\right]\ket{g,0}+ e^{\ii \left(\varphi _1-\varphi _2\right)} \sin \left(\frac{t_1}{2}\right) \left(e^{\ii \left(\theta +\varphi _2\right)} \cos \left(\frac{t_2}{\sqrt{2}}\right) \cos \left(\frac{t_R}{\sqrt{2}}\right)+\sin \left(\frac{t_2}{\sqrt{2}}\right) e^{\ii \phi _1} \sin
   \left(\frac{t_R}{\sqrt{2}}\right)\right)\ket{e,1}\nonumber\\
  &+e^{\ii \varphi _1} \sin \left(\frac{t_1}{2}\right) \left(\cos \left(\frac{t_2}{\sqrt{2}}\right) e^{\ii \left(\theta -\varphi_1\right)} \sin \left(\frac{t_R}{\sqrt{2}}\right)-e^{-\ii \varphi _2} \sin \left(\frac{t_2}{\sqrt{2}}\right) \cos
   \left(\frac{t_R}{\sqrt{2}}\right)\right)\ket{g,2}.
 \end{align}
% Here, the modulation is arising due to the interference between $\ket{e,1}$ and $\ket{g,2}$.  Here,  $\ket{\Psi}_{B \vert R}$, and $\ket{\Psi}_{BR \vert R}$  are of similar forms with superposition of $\ket{g,0}$, $\ket{e,1}$ and $\ket{g,2}$. However, only $\ket{\Psi}_{BR \vert R}$ makes modulation by the phase of the verification pulse. Such an effect occurs due to the non-commutativity of red pulses. Similar effect can happen for blue pulses due to their non-commutativity.
% The maximum interference occurs at $t_1=\pi$, $t_2=\tfrac{\pi}{2\sqrt{2}}$ and $t'_1=\tfrac{\pi}{2\sqrt{2}}$. With these parameters, the energy of the state in (\ref{eq:PsiBR})  is given as $3/2$.
% After blue pulse for verification, the state evolves to 
%The modulation arises from the interference between $\ket{e,1}$ and $\ket{g,2}$. Both $\ket{\Psi}_{B \vert R}$ and $\ket{\Psi}_{BR \vert R}$ contain superpositions of $\ket{g,0}$, $\ket{e,1}$, and $\ket{g,2}$, but only $\ket{\Psi}_{BR \vert R}$ shows modulation affected by the verification pulse phase. This is due to the non-commutativity of red pulses, with a similar effect possible for blue pulses. 
%Maximum interference occurs between $\ket{g,2}$ and $\ket{e,1}$ at $t_1 = \pi$, $t_2 = \tfrac{\pi}{2\sqrt{2}}$, and $t_1' = \tfrac{\pi}{2\sqrt{2}}$, where the energy of the state (\ref{eq:PsiBR}) is $3/2$,  when the vacuum is unoccupied due to complete population transfer by the first preparation pulse. If the vacuum remains occupied, interference is smaller.
{Maximum interference for this state occurs when the verification pulse couples the dominant motional components, in this case $\ket{g,2}$ and $\ket{e,1}$. 
The magnitude of this interference is reduced if the initial state preparation is imperfect (e.g., if population remains in the vacuum state).}

% With decoherence factor $w$, the modulation of $P_g$ is given as:
% \begin{align}
%     P_g=\frac{1}{4} \left(2 \sin ^2\left(\frac{t_1}{2}\right) \left(-(w-1) \sin \left(\sqrt{2} t_2\right) \sin \left(\sqrt{2} t'_1\right) \cos \left(\phi _2-\phi'_1\right)-\cos \left(\sqrt{2} t_2\right) \cos
%    \left(\sqrt{2} t'_1\right)\right)+\cos \left(t_1\right)+3\right).
% \end{align}
% Again, only the modulation amplitude depends on $w$. Again from known $P(n)$, we can specify $t'_1$ by looking at the non-modulating value, and in turn  specify $w$.
With decoherence factor $w$, $P_g$ modulation by $\phi_1$:
\begin{align}
    P_g=\frac{1}{4} \left(2 \sin ^2\left(\frac{t_1}{2}\right) \left(-(w-1) \sin \left(\sqrt{2} t_2\right) \sin \left(\sqrt{2} t_R\right) \cos \left(\varphi _2-\phi_1\right)-\cos \left(\sqrt{2} t_2\right) \cos
   \left(\sqrt{2} t_R\right)\right)+\cos \left(t_1\right)+3\right).
\end{align}
Again, only modulation amplitude depends linearly on $w$. From known $P(n)$,  $t'_1$ can be precisely specified from the non-modulating value, and thus $w$ precisely determined.

After blue verification  pulse (instead of red), the state evolves to
 \begin{align}
  \ket{\Psi}_{BR\vert B}=&\Big(\cos \left(\frac{t_1}{2}\right) \cos \left(\frac{t_B}{2}\right)+\sin \left(\frac{t_1}{2}\right) \cos \left(\frac{t_2}{\sqrt{2}}\right) e^{\ii \left(\varphi _1+\theta-\phi _1\right)}
   \sin \left(\frac{t_B}{2}\right)\Big)\ket{g,0}\nonumber\\
   &-e^{\ii \left(\varphi _1-\varphi _2\right)} \sin \left(\frac{t_1}{2}\right) \sin \left(\frac{t_2}{\sqrt{2}}\right) \cos \left(\frac{1}{2} \sqrt{3} t_B\right)\ket{g,2}\nonumber\\
   &+\Big(e^{\ii (\varphi _1+\theta)} \sin \left(\frac{t_1}{2}\right) \cos \left(\frac{t_2}{\sqrt{2}}\right) \cos \left(\frac{t_B}{2}\right)-\cos \left(\frac{t_1}{2}\right) e^{\ii \phi_1} \sin
   \left(\frac{t_B}{2}\right)\Big)\ket{e,1}\nonumber\\
   &+\sin \left(\frac{t_1}{2}\right) \sin \left(\frac{t_2}{\sqrt{2}}\right) e^{\ii \left(\phi_1+\varphi _1-\varphi _2\right)} \sin \left(\frac{1}{2} \sqrt{3} t_B\right) \ket{e,3}.
 \end{align}
Here, the last pulse exhibits the interference of  $\ket{g,0}\leftrightarrow \ket{e,1}$. 
In both cases, we again see the equivalence  of varying $\phi_1$ and $\theta$.

\section{Effect of sideband pulses}
\label{Appendix:pulse}
The effect of the \(j\)-th red pulse (\(l = -1\)) or blue pulse (\(l = 1\)) is described by a unitary transformation \(U_l^{(j)}\), as detailed in \cite{Leibfried2003}:
\begin{align}
    U_l^{(j)}\ket{\Psi}^{(j-1)} &= \sum_{n=\text{even}} \Bigg(c_{l+n}^{(j-1)} \cos \left(\frac{t \Omega_{n,n+l}}{2}\right) - \ii e^{\ii \left(\phi + \frac{\pi}{2}\right)} c_n^{(j-1)} \sin \left(\frac{t \Omega_{n,n+l}}{2}\right)\Bigg) \ket{e,n+l} \nonumber \\
    &+ \Bigg(c_n^{(j-1)} \cos \left(\frac{t \Omega_{n,n+l}}{2}\right) - \ii e^{\ii \left(\phi + \frac{\pi}{2}\right)} c_{l+n}^{(j-1)} \sin \left(\frac{t \Omega_{n,n+l}}{2}\right)\Bigg) \ket{g,n},
    \label{eq:leibfried}
\end{align}
where \(\eta\) is the Lamb-Dicke parameter, and \(\Omega_{n,n+l} = \eta \Omega_0 \sqrt{n + \frac{l+1}{2}} \) is the Rabi frequency, with \(\Omega_0\) being the fundamental Rabi frequency.
% The coefficients \( c_n^{(j-1)} \) are determined by the preceding \( j-1 \) pulses. This transformation occurs by all pulses, including those used for verification.
% Full (half) of the population transfers from the lower Fock state to the higher Fock state at $\frac{t \eta \Omega_{n,n+l}}{2} = \frac{\pi}{2}$ ($\tfrac{\pi}{4}$). 
% In the experiment, to simplify control, we set the strength fixed to $\frac{t \eta\Omega_{0}}{2} = \frac{\pi}{4}$. Below, we set $\eta \Omega_0=1$ for simplicity. %The smallest number of preparation pulses for $\ket{0}\bra{3}$ coherence to be prepared is 3.
% As in Fig. \ref{fig:macroscopicitynbarvsN}, this pulse length rapidly increases the macroscopicity of the prepared state.
The coefficients $ c_n^{(j-1)} $ are determined by the preceding $ j-1 $ pulses. This transformation applies to all pulses, including those used for verification.
The pulse area, the integrated area under the applied pulse envelope, is a function of both the pulse strength (amplitude) and the pulse duration (length). 
We assume the Rabi frequency to be time-independent.

\section{Effect of decoherence and determination of the coherence parameter $w$}
\label{Appendix:decoherence}

A physically motivated model for decoherence assumes that off-diagonal terms in the density matrix, $|n\rangle\langle m|$, decay at a rate dependent on their separation $k = |n-m|$. 
This captures that long-range coherences are more fragile. 
Under a model of independent, local dephasing events, the coherence of each term in the density matrix of the ideal prepared state, $\rho^{(\text{ideal})}$, would be suppressed exponentially:
\begin{equation}
\rho_{nm} = \rho_{nm}^{(\text{ideal})} w^{|n-m|},
\label{eq:wk_model}
\end{equation}
where the parameter $w \le 1$ represents the fundamental coherence factor between adjacent Fock states.

However, as detailed in Appendix \ref{app:offdiag}, our two-pulse verification sequence is primarily sensitive only to short-range coherences where $|n-m| \le 3$. For a state with a high degree of overall coherence ($w \approx 1$), the suppression factors for these relevant terms ($w$, $w^2$, and $w^3$) are all numerically close. Therefore, for the practical analysis in this work, we can simplify this picture by adopting a single \textit{effective} coherence parameter, which we also denote as $w$. This single parameter effectively captures the average suppression of the specific off-diagonal terms that our measurement probes.

This simplification allows us to model the state with a single parameter that directly relates to the observed interference contrast. The partially coherent state is written as an incoherent mixture of the ideal pure state $|\Psi\rangle$ and its fully dephased version:
\begin{equation}
\rho[w] = w |\Psi\rangle \langle\Psi|+(1-w) \text{diag}[|\Psi\rangle \langle\Psi|],
\label{eq:effective_model}
\end{equation}
where $\text{diag}[\rho]$ is the matrix containing only the diagonal elements of $\rho$. The modulation amplitude in our interference measurements is then suppressed by this effective factor $w$ relative to the ideal coherent case ($w=1$). This type of state can result from the combined effect of qubit dephasing and oscillator dephasing, modeled here phenomenologically. The other type of state model with diminished mutual information is
\begin{equation}
\rho[w] = w |\Psi\rangle \langle\Psi| + (1 - w)(\frac{1}{2} |e\rangle \langle e| \otimes |\psi_{-}\rangle \langle\psi_{-}| + \frac{1}{2} |g\rangle \langle g| \otimes |\psi_{+}\rangle \langle\psi_{+}|),
\label{eq:incoherentDMS}
\end{equation}
where $|\psi_{\pm}\rangle$ are even/odd states associated to each qubit basis.
Dominant physical mechanisms contributing to the reduction of \(w\) (i.e., an increase in the incoherent fraction \(1-w\)) in our trapped ion system  include motional heating from noisy trap potentials or patch potentials \cite{Brownnutt2015RevModPhys.87.1419}, uncompensated slow laser phase/frequency noise, and magnetic field fluctuations that cause shot-to-shot variations in the prepared state's relative phase \(\theta\). For the macroscopic motional states explored, motional heating and such environmental phase drifts are expected to be primary contributors.
For the full decoherence models, we should use the descriptions as in \cite{TurchettePRA2000Decoherence}.

Similar suppression of the verification pulse phase modulation contrast can also occur if the Ramsey pulse length $t_{R,B}$ is small. %, as shown in Fig. \ref{fig:modulationcurve}. 
To exclude these effects, we %need to perform a Rabi flop to check the Fock weight distribution $P(n)$ right before and after the Ramsey pulse. Knowing the phonon number distribution $P(n)$ of the output state allows us to determine whether the reduced modulation is due to insufficient pulse length or decoherence. 
employ a multi-step analysis. First, the Fock distribution $P(n)$ measured via Rabi flopping (Sec II.c) constrains the prepared state amplitudes $|c_n|$. Second, the average value of $P_g$ over the verification phase scans depends on $|c_n|^2$ and the verification pulse lengths $t_{R,B}$ (via terms like Eq. \ref{eq:nonmodulation} and its two-pulse extension), allowing $t_{R,B}$ to be estimated. Finally, the amplitude of the observed $P_g$ modulation (dependent on terms like Eq. \ref{eq:modulationamplitude}) is compared to the amplitude expected for the estimated $c_n$ and $t_{R,B}$ in the ideal coherent case ($w=1$), allowing the coherence factor $w$ to be determined. 
This disentanglement of \(w\) from errors in \(t_{R,B}\) relies on \(t_{R,B}\) primarily affecting the average population transfer (DC offset of \(P_g\)) and overall scaling of modulation, while \(w\) specifically scales the relative amplitude of the coherent interference terms. The robustness of this separation can be challenged if decoherence during preparation significantly alters \(P(n)\) itself, as this affects the \(|c_n|\) used to estimate \(t_{R,B}\), potentially leading to propagated errors. However, the distinct mathematical dependencies generally allow for their separate estimation, assuming errors in \(t_{R,B}\) do not systematically mimic the phase-dependent signatures of coherence loss captured by \(w\). A brief sensitivity analysis indicates that a typical 1\% mismatch in the driving field strength (affecting the Rabi frequency \(\Omega\)) and thus the verification pulse area \(t_{R,B}\), cause the subsequent impact on the estimated coherent fraction \(w\) is correspondingly small by approximately 1-2\%, supporting the robustness of this disentanglement under typical experimental imperfections.

Consider a prepared state in a partially coherent mixture:
\begin{align}
\rho_\Psi[w] \equiv w\ket{\Psi}\bra{\Psi} + (1-w)\operatorname{diag}[\ket{\Psi}\bra{\Psi}],
\label{eq:partialcoherent}
\end{align}
where $\ket{\Psi} = \sum_n c_n  \ket{f_n,n}$ (real coefficients $c_n=|c_n| e^{\ii \phi_n}$, and {$|f_n\rangle$} can be either {$|e\rangle$ or $|g\rangle$} depending on the number parity of $n$).
%$\operatorname{diag}[\rho]$ denotes the diagonal matrix of $\rho$
The modulation amplitude is suppressed by $w$, and so are the contrast and the visibility, for a general state $\rho_\Psi[w]$.
% Contrast is sensitive to the overall coherence of the state, $w$.
% High visibility ($V \approx 1$) is a robust signature that the interfering pathways are perfectly balanced by our verification protocol.
For a red sideband verification pulse, the non-modulating term of $P_g$ (w.r.t. pulse phase) is:
\begin{align}
    \sum_{n=\text{even}} |c_n|^2 \cos^2\left(\frac{\sqrt{n/2}}{\sqrt{2}} t_R\right) + \sum_{n=\text{odd}} |c_n|^2 \sin^2\left(\frac{\sqrt{(n+1)/2}}{\sqrt{2}} t_R\right).
    \label{eq:nonmodulation}
\end{align}
The non-modulating term Eq. (\ref{eq:nonmodulation}) can be predicted solely from the measured populations $P(n) \approx |c_n|^2$. 
Modulation amplitude is:
\begin{align}
    w \left| \sum_{n=1} c_{2n-1} c_{2n} e^{\ii (\varphi_{2n} - \varphi_{2n-1})} \sin\left(\sqrt{2n} t_R\right) \right|.
    \label{eq:modulationamplitude}
\end{align}
The modulation amplitude Eq. (\ref{eq:modulationamplitude}) is then measured, and by comparing it to the ideal case ($w=1$), we can extract the coherence factor $w$ for the prepared state.
%These equations (\ref{eq:nonmodulation}) and (\ref{eq:modulationamplitude}) formalize our methodology for an arbitrary number of pulses. 
The above derivation shows that the amplitude of the measured interference fringes is directly proportional to $w$, making it a direct experimental signature of the state's quantum coherence.

%Now using the contrast by second verification pulse phase $\phi'[2]$ for various first verification pulse phase  $\phi'[1]$, we can precisely determine the decoherence factor $w$ in the prepared setup.
The values of  the coherent fraction \(w\) that explain the experimental data best using this procedure are given as: For a setup of $N=1,2,3,4,6$, $1-w\approx0.05$, $0.04-0.12$, $0.083$, $0.137-0.143$ and $0.08-0.13$.
%For longer preparation pulse sequences, it gets harder to estimate $w$ due to the existence of many parameters.
%For $N=6$, it is estimated to be $w\approx0.08-0.13$. 
For longer preparation pulse sequences, determination of $w$ has more uncertainty due to the existence of many parameters {and potential propagation of errors in the estimation}. For $N=8,10,12$, the incoherence fraction  is about $1-w\approx0.08-0.13, 0.09-0.2, 0.08-0.13$.
 {These values indicate a generally high degree of coherence.
 {The observed complex scaling of $1-w$ with the number of pulses $N$ suggests an interplay of multiple physical mechanisms. 
 Plausible contributions include time-dependent effects such as motional heating and qubit dephasing, which accumulate over the sequence duration, and pulse-dependent errors from laser imperfections that accumulate with each operation. 
 The non-trivial scaling likely arises because the growing macroscopic state becomes differentially sensitive to these error channels. 
 A detailed analysis to disentangle these contributions would be a valuable next step but is beyond the scope of this work.}
 %An observable simple scaling trend for \(1-w\) (e.g., linear or exponential with total pulse time or \(N\)) is only weakly apparent from the data, suggesting that the interplay of decoherence mechanisms is complex and does not follow a straightforward model for the explored range of \(N\). 
 This complexity influences the ultimate scalability of preparing highly coherent macroscopic states.}

\subparagraph{Verification of  off-diagonal elements}

% As the {verification process} and measurement can be described by the sum of all the effects of individual density matrix elements, we can alternatively look to how individual off-diagonal density matrix elements such as $\ket{n}\bra{n-k}$ for non-zero $k$ evolve  and contribute to the measurement. {Appendix \ref{app:offdiag} details how specific off-diagonal terms like $\ket{n}\bra{m}$ in the prepared state contribute to $P_g$ modulation at distinct frequencies dependent on the verification phases ($\phi'_1, \phi'_2$), enabling their detection.} 
{Alternatively, we can analyze the coherence structure in the frequency domain. 
Each off-diagonal element of the prepared state's density matrix, such as $|n\rangle\langle m|$, contributes to the modulation of $P_g$ at a unique frequency with respect to the verification phases $\phi_1$ and $\phi_2$. 
This allows us to use Fourier analysis of the measured $P_g(\phi_1, \phi_2)$ signal to identify the presence and magnitude of specific coherence terms.}
In order for them to have measurable effects, the %\cmmnt{diagonal elements in the density matrix should be generated by the verification pulses}
{verification pulses must coherently couple the relevant states before measurement}.

Experimentally, we can observe such effects by analyzing the frequency components of the $P_g$ modulation as a function of $\phi_1$ and $\phi_2$. 
% For example %if the inverse sequence is made of 2 pulses, when we vary their phases, the prepared $\ket{0}\bra{1}$ element has effect on the output $\ket{0}\bra{0}$, $\ket{1}\bra{1}$  and $\ket{2}\bra{2}$ modulating at $e^{\pm\ii \phi[1]}$ where $\phi[1]$ is a blue sideband pulse. In a similar way, the prepared $\ket{0}\bra{2}$ element has effect on the output  $\ket{1}\bra{1}$  and $\ket{2}\bra{2}$ modulating at $e^{\pm(\ii \phi[1]-\ii \phi[2])}$. In comparison, the prepared $\ket{0}\bra{3}$ element has effect on the output  $\ket{1}\bra{1}$  and $\ket{2}\bra{2}$ modulating at $e^{\pm(\ii 2\phi[1]-\ii \phi[2])}$. 
% {, Appendix E shows that} the prepared $\ket{e,0}\bra{g,3}$ element {can contribute to $P_g$ terms} modulating at frequencies like $e^{\pm\ii (2\phi'_1-\phi'_2)}$ {when using a RSB($\phi'_1$) followed by a BSB($\phi'_2$) for verification}.
% This $k=3$ %\cmmnt{is the largest off-diagonal element we can directly evidence the existence in experiment using two sequential verification pulses.}
% represents coherence between states differing by 3 phonons. 
For example, as detailed in Appendix \ref{appendix:longer}, a coherence term like $|g,0\rangle\langle e,3|$ (representing coherence between states differing by 3 phonons) will uniquely generate a modulation signal at a frequency proportional to $2\phi_1 + \phi_2$ under an RSB-BSB verification sequence. 
Observing a signal at this specific frequency is therefore direct evidence of this ``deep" coherence element.
While potentially challenging to measure due to small coefficients, observing modulation at the corresponding frequency provides {a signature consistent with} such deep coherence.
Therefore, using the {characteristic modulation frequencies}, we can distinguish the effects and {signatures} of the prepared off-diagonal elements.    
%However, such a deep coherence cannot be identified if only carrier and one sideband pulse is used sequentially.

By performing a Fourier analysis of the measured $P_g(\phi_1, \phi_2)$ data, we can identify signatures of specific off-diagonal coherence terms. 
For the $N=3$ state, we detect a non-zero modulation amplitude at the frequency corresponding to the $|g,0\rangle\langle e,3|$ coherence. 
The fitted amplitude of this signal is 0.105, in close agreement with the ideal theoretical prediction of 0.095, confirming the presence of this long-range coherence. 
Similarly, for $N=4$, we detect a small but non-zero signature for the $|e,1\rangle\langle g,4|$ coherence. 
The existence of these specific, higher-order coherence terms provides definitive, quantitative evidence of the complex entangled structure created by our method.
The detection of these non-zero Fourier coefficients is a direct measurement of higher-order coherence within the prepared state.

\section{Derivation of the orthogonal mapping between coherence and probe phases}
\label{appendix:equivalnece}

% The commutation relation of the phase rotation operators and evolution operator by sideband pulses can show the equivalence of the varying pulse phases and the coherence by phase rotations. 
% For example, by the qubit rotation operator we have the transformation of red sideband interaction
% \begin{align}
%     \exp[-\ii \frac{\theta}{2}\sigma_z ]\exp[\ii \kappa (\sigma_+ \hat{a}+\sigma_- \hat{a}^\dagger)]\exp[\ii \frac{\theta}{2}\sigma_z ]=\exp[\ii \kappa (\sigma_+ \hat{a}e^{-\ii \theta}+\sigma_- \hat{a}^\dagger e^{\ii \theta})].
% \end{align}
% Similarly, 
% \begin{align}
%     \exp[-\ii \theta\hat{n} ]\exp[\ii \kappa (\sigma_+ \hat{a}+\sigma_- \hat{a}^\dagger)]\exp[\ii \theta \hat{n}  ]=\exp[\ii \kappa (\sigma_+ \hat{a}e^{\ii \theta}+\sigma_- \hat{a}^\dagger e^{-\ii \theta})].
% \end{align}
% Similar relationships are obtained for blue sideband pulses.
% This shows that the varying $\theta$ is effectively equivalent to varying the phase of the pulses, and thus we do not need active phase rotations.

This section provides the mathematical derivation for the measurement method used in the main text. We demonstrate how the coherence of the prepared state $|\Psi\rangle$, parameterized by the qubit-oscillator phase $\theta$ and the internal oscillator phase $\Theta$, can be independently measured by controlling the external phases $\phi_1$ and $\phi_2$ of the verification pulses. We show that the commutation relations of the phase rotation and sideband interaction operators create an orthogonal mapping, where the sum of the probe phases $\frac{\phi_1 + \phi_2}{2}$ addresses $\theta$, and the difference $\frac{\phi_1 - \phi_2}{2}$ addresses $\Theta$.

The commutation relations between phase rotation operators and the evolution operator induced by sideband pulses illustrate the {mapping between internal coherence and the phase of verification pulses.} % equivalence between varying pulse phases and coherence through phase rotations.
Specifically, the transformation under the qubit rotation operator for the red sideband interaction is given by:
\begin{align}
\exp\left[\ii \kappa (\sigma_+ \hat{a}e^{-\ii \phi}  + \sigma_- \hat{a}^\dagger e^{\ii \phi} )\right] \exp\left[\ii \frac{\theta}{2} \sigma_z \right] = \exp\left[\ii \frac{\theta}{2} \sigma_z \right] \exp\left[\ii \kappa (\sigma_+ \hat{a} e^{-\ii (\overbrace{\theta+\phi}^{\phi(\theta)})} + \sigma_- \hat{a}^\dagger e^{\ii (\theta+\phi)})\right].
\label{eq:thetaequiv1}
\end{align}
{This shows the effect of a qubit phase rotation on the red sideband (RSB) interaction. Physically, this demonstrates that applying a qubit rotation corresponding to phase $\theta$ \textit{before} the interaction is mathematically identical to shifting the external laser phase $\phi$ by $+\theta$ \textit{during} the interaction.}

Similarly, the adjoint relation for oscillator coherence by oscillator phase rotation operation:
\begin{align}
\exp\left[\ii \kappa (\sigma_+ \hat{a} e^{-\ii \phi}+ \sigma_- \hat{a}^\dagger e^{\ii \phi})\right] \exp\left[\ii \Theta \hat{n} \right] = \exp\left[\ii \Theta \hat{n} \right] \exp\left[\ii \kappa (\sigma_+ \hat{a} e^{-\ii (\overbrace{-\Theta+\phi}^{\phi(\Theta)})} + \sigma_- \hat{a}^\dagger e^{\ii (-\Theta+\phi)})\right].
\end{align}
{Conversely, this equation  shows the effect of a motional phase rotation. This transformation is equivalent to shifting the external laser phase $\phi$ by $-\Theta$. The opposite sign of this phase shift compared to the qubit rotation is the fundamental reason our method can distinguish the two types of coherence.}

Analogous transformations apply to blue sideband pulses. 
Explicitly, 
\begin{align}
\exp\left[\ii \kappa (\sigma_+ \hat{a}^\dagger e^{-\ii \phi'}  + \sigma_- \hat{a} e^{\ii \phi'} )\right] \exp\left[\ii \frac{\theta}{2} \sigma_z \right] = \exp\left[\ii \frac{\theta}{2} \sigma_z \right] \exp\left[\ii \kappa (\sigma_+ \hat{a}^\dagger e^{-\ii (\overbrace{\theta+\phi'}^{\phi'(\theta)})} + \sigma_- \hat{a} e^{\ii (\theta+\phi')})\right].
\end{align}
Similarly, the adjoint relation for oscillator coherence by oscillator phase rotation operation:
\begin{align}
 \exp\left[\ii \kappa (\sigma_+ \hat{a}^\dagger e^{-\ii \phi'}+ \sigma_- \hat{a} e^{\ii \phi'})\right] \exp\left[\ii \Theta \hat{n} \right] = \exp\left[\ii \Theta \hat{n} \right]\exp\left[\ii \kappa (\sigma_+ \hat{a}^\dagger e^{-\ii (\overbrace{\Theta+\phi'}^{\phi'(\Theta)})} + \sigma_- \hat{a} e^{\ii (\Theta+\phi')})\right].
\label{eq:thetaequiv4}
\end{align}
{These equations  show the analogous transformations for a blue sideband (BSB) pulse. Crucially, for a BSB interaction, both the qubit rotation ($\theta$) and the motional rotation ($\Theta$) result in a shift of the \textit{same sign} ($+\theta$ and $+\Theta$) in the external phase $\phi'$. The difference in how the motional phase $\Theta$ affects RSB versus BSB pulses is what allows for the orthogonal mapping.}
% These equations show that varying \(\theta\) and $\Theta$ {shows equivalent effect in the interference} to changing the phases $\phi, \phi'$ of the pulses for the scan, eliminating the need for active phase rotations other than the sideband pulses interactions with varied phases.
% For oscillator coherence verification, we therefore need simultaneous scan of both pulses in the verification pulses.
% Especially, the phase rotates in opposite directions for blue and red sideband pulses phases by oscillator phase rotation.
{These operator identities demonstrate that the internal coherence phases of the state ($\theta, \Theta$) and the external control phases of the verification pulses ($\phi_1, \phi_2$) are interchangeable. When we apply a two-pulse verification sequence (e.g., RSB with phase $\phi_1$ followed by BSB with phase $\phi_2$), the total phase dependence of the final state arises from combinations of these shifts. The qubit-oscillator coherence $\theta$ will manifest in terms dependent on $(\phi_1 + \phi_2)/2$ because its effect has the same sign for both pulses. In contrast, the internal oscillator coherence $\Theta$ will manifest in terms dependent on $(\phi_1 - \phi_2)/2$ because its effect has the opposite sign for RSB versus BSB pulses.}

Consequently, the probability of qubit ground state $P_g$ from the prepared state $\sum_{j=0}^N c_j \ket{f(j),j}$ where $f(j)=e,g$ depending on the parity of Fock number $j$ after the verification pulses {(e.g., RSB($\phi_1$) then BSB($\phi_2$)) acting on the prepared state $\ket{\Psi}$ can be written schematically} as
\begin{align}
    P_g &= \text{const} + \sum_{k,l} A_{kl} \cos(k\phi_1 + l\phi_2) + B_{kl} \sin(k\phi_1 + l\phi_2) 
    % P_g&=const+F(c_1c_2,c_3c_4,...)\cos[\phi'_1] +G(c_0c_1,c_2c_3,...)\cos[\phi'_2]+H(c_0c_2,c_1c_3,c_2c_4,...)\cos[\phi'_1-\phi'_2]\nonumber\\
    % &+I(c_1c_4,c_3c_6,...)\cos[2\phi'_1-\phi'_2]\nonumber\\
    % &+F'(c_1c_2,c_3c_4,...)\sin[\phi'_1] +G'(c_0c_1,c_2c_3,...)\sin[\phi'_2]+H'(c_0c_2,c_1c_3,c_2c_4,...)\sin[\phi'_1-\phi'_2]\nonumber\\
    % &+I'(c_1c_4,c_3c_6,...)\sin[2\phi'_1-\phi'_2].
\end{align}
{The amplitudes of the oscillating terms, $A_{kl}$ and $B_{kl}$, are directly proportional to the magnitude of the coherence in the prepared state, while the phase offsets of the fringes reveal the internal phases $\theta$ and $\Theta$.}
%\cmmnt{The contrast of $P_g$ can be calculated analytically as a function of $F,G,H,I$. Now this form can be simply converted to the qubit phase $\phi=\phi'_1+\phi_2$ and oscillator phase $\theta=\phi'_1-\phi_2$, whose equivalence was noted in (\ref{eq:thetaequiv1}-\ref{eq:thetaequiv4}).}  
Under the partial coherence model (Eq. \ref{eq:partialcoherent}), this functional form is still maintained, while the amplitudes of the oscillating terms ($A_{kl}, B_{kl}$) are linearly diminishing by $w$.
% The contrast of $P_g$ can be calculated analytically as a function of $F,G,H,I$.
% Now this form can be simply converted to the qubit phase $\phi=\phi'_1+\phi_2$ and oscillator phase $\theta=\phi'_1-\phi_2$, whose equivalence was noted in (\ref{eq:thetaequiv1}-\ref{eq:thetaequiv4}).

% {Under decoherence model delineated in Appendix \ref{Appendix:decoherence}, this functional form is still maintained, while $F,G,H,I$ are linearly diminishing by $w$.}

Applying a second verification pulse allows for a more detailed analysis. 
% The effect of the verification pulses is summarized in Appendix \ref{app:offdiag}.
% The POVM of the measurement setup (i.e. qubit detection) are sum of the three terms as $\hat{\Pi}=\hat{\Pi}_1+\hat{\Pi}_2+\hat{\Pi}_3$ and $1-\hat{\Pi}$, where:
To be more concrete, we can explicitly write down the Positive Operator-Valued Measure (POVM) element $\Pi$ corresponding to a final measurement of the qubit in the ground state $|g\rangle$. For an RSB($\phi_1$)-BSB($\phi_2$) verification sequence, this operator can be decomposed into three terms, $\Pi = \Pi_1 + \Pi_2 + \Pi_3$, which describe the different physical processes contributing to the final $P_g$.    
\begin{align}
\hat{\Pi}_{3} =\sum_{k=1} -\frac{1}{2}\ket{e}\bra{g}\otimes \ket{2k-1}\bra{2(k+1)} \exp\left[\ii (2\phi_1 - \phi_2)\right] & \sin\left(\sqrt{2k+1}  t_2'\right)  \sin\left(\sqrt{\frac{k}{2}}  t_1'\right) \sin\left(\sqrt{\frac{k+1}{2}}  t_1'\right),
\end{align}
This term contains the operator $|e\rangle\langle g|$ and describes transitions between the two qubit states. It is responsible for probing the \textit{qubit-oscillator coherence of the third order Fock coherence}.
{For $t_2'=0$, this term  reduces to $0$.}

\begin{align}
\hat{\Pi}_2& =\ket{g}\bra{g}\otimes\sum_{k=1} -\frac{1}{2} \ket{2(k-1)}\bra{2(k+1)} \exp\left[\ii(\phi_1 - \phi_2)\right] & \sin\left(\sqrt{2k-1}  t_2'\right)  \cos\left(\sqrt{\frac{k-1}{2}}  t_1'\right) \sin\left(\sqrt{\frac{k}{2}}  t_1'\right)\nonumber\\
&+\ket{e}\bra{e}\otimes\sum_{k=1} \frac{1}{2} \ket{2k-1}\bra{2k+1}\exp\left[\ii(\phi_1 - \phi_2)\right] & \sin\left(\sqrt{2k+1}  t_2'\right)   \cos\left(\sqrt{\frac{k+1}{2}}  t_1'\right) \sin\left(\sqrt{\frac{k}{2}}  t_1'\right).
\end{align}
This term contains operators like $|g\rangle\langle g|$ and $|e\rangle\langle e|$, describing transitions within the same qubit subspace. It therefore probes the \textit{internal oscillator coherence} within each parity subspace. 
{For $t_2'=0$, this term is reduced to $0$ as well.
This term remains constant for $\phi_1 - \phi_2=const$.}

\begin{align}
\hat{\Pi}_1&  =\ket{g}\bra{e}\otimes\sum_{k=1} \frac{1}{2} \ket{2k-2}\bra{2k-1} \exp\left[-\ii\phi_2\right] & \sin\left(\sqrt{2k-1}  t'_2\right)  \cos\left(\sqrt{\frac{k-1}{2}}  t'_1\right) \cos\left(\sqrt{\frac{k}{2}}  t'_1\right)\nonumber\\
&-\ket{e}\bra{g}\otimes\sum_{k=1}\frac{1}{2} \ket{2k-1}\bra{2k} \exp\left[\ii\phi_1\right] & \sin\left(\sqrt{2k}  t'_1\right)  \times \left[ \sin^2\left(\frac{\sqrt{2k-1}}{2} t'_2\right) - \cos^2\left(\frac{\sqrt{2k+1}}{2} t'_2\right) \right].
\end{align}
This term also contributes to the interference signal, representing cross-terms in the measurement. This term can be  detected by the single-pulse measurement, i.e. when $t_2'=0$.
{It then reduces to 
\begin{align}
\hat{\Pi}_1=-\ket{e}\bra{g}\otimes\sum_{k=1}\frac{1}{2} \ket{2k-1}\bra{2k} \exp\left[\ii\phi'_1\right] & \sin\left(\sqrt{2k}  t'_1\right).
\end{align}
}

These are substituted directly into the following probabilities.
The expression for the term containing $\exp[3\ii  \Theta]$ is given by:
\begin{align}
P_g^{(3)} =\sum_{k=1} -\frac{1}{2}c_{2(k+1)}c^*_{2k-1} \exp\left[\ii (2\phi_1' - \phi_2')\right] & \sin\left(\sqrt{2k+1}  t_2'\right)  \sin\left(\sqrt{\frac{k}{2}}  t_1'\right) \sin\left(\sqrt{\frac{k+1}{2}}  t_1'\right).
\end{align}
The expression for the term containing $\exp[2\ii  \Theta]$ is given by:
\begin{align}
P_g^{(2)}& =\sum_{k=1} -\frac{1}{2} c_{2(k+1)}c^*_{2(k-1)} \exp\left[\ii(\phi_1' - \phi_2')\right] & \sin\left(\sqrt{2k-1}  t_2'\right)  \cos\left(\sqrt{\frac{k-1}{2}}  t_1'\right) \sin\left(\sqrt{\frac{k}{2}}  t_1'\right)\nonumber\\
&+ \frac{1}{2} c_{2k+1}c^*_{2k-1}\exp\left[\ii(\phi_1' - \phi_2')\right] & \sin\left(\sqrt{2k+1}  t_2'\right)   \cos\left(\sqrt{\frac{k+1}{2}}  t_1'\right) \sin\left(\sqrt{\frac{k}{2}}  t_1'\right).
\end{align}
The expression for the term containing $\exp[\ii  \Theta]$ is given by:
\begin{align}
P_g^{(1)}&  =\sum_{k=1} \frac{1}{2} c_{2k-1}c_{2k-2}^* \exp\left[-\ii\phi'_2\right] & \sin\left(\sqrt{2k-1}  t'_2\right)  \cos\left(\sqrt{\frac{k-1}{2}}  t'_1\right) \cos\left(\sqrt{\frac{k}{2}}  t'_1\right)\nonumber\\
&-\frac{1}{2} c_{2k}c_{2k-1}^* \exp\left[\ii\phi'_1\right] & \sin\left(\sqrt{2k}  t'_1\right)  \times \left[ \sin^2\left(\frac{\sqrt{2k-1}}{2} t'_2\right) - \cos^2\left(\frac{\sqrt{2k+1}}{2} t'_2\right) \right].
\end{align}
{The terms depending on $\exp[-\ii k  \Theta]$ for $k=1,2,3$ is given simply by their complex conjugates.}
The non-modulating (DC) component of $P_g$ for a two-pulse sequence is given by:
\begin{align}
    &\sum_{n=even} |c_n|^2 \Big(\cos[\frac{\sqrt{n+1}}{2}t'_2]\cos[\frac{\sqrt{n}}{2}t'_1]^2+\sin[\frac{\sqrt{n-1}}{2}t'_2]^2\sin[\frac{\sqrt{n}}{2}t'_1]^2\Big)\nonumber\\
    &+ \sum_{n=odd} |c_n|^2 \Big(\cos[\frac{\sqrt{n+2}}{2}t'_2]\sin[\frac{\sqrt{n+1}}{2}t'_1]^2+\sin[\frac{\sqrt{n}}{2}t'_2]^2\cos[\frac{\sqrt{n+1}}{2}t'_1]^2\Big).
\end{align}
%Again, using this term, we can determine $t'_1$ and $t'_2$, and then estimate $w$ from modulation contrast.
As before, this DC term depends on the measured populations and the known verification pulse area ($t'_1, t'_2$), while the modulation amplitude depends linearly on the coherence factor $w$, allowing for its robust estimation.
Non-zero co-existence of $P_g^{(1)}$, $P_g^{(2)}$ and $P_g^{(3)}$ signifies the complex structure in Fock basis, minimally containing terms of $\{\ket{0},\ket{1},\ket{3}\}$, which can be directly measured by the Fourier series expansion of the total probability $P_g$. 
By varying verification pulse area ($t'_1, t'_2$), it is also possible to explore individual coherence of $\ket{n}\bra{m}$ for different $n,m$.
%{The modulation by }

\begin{figure}[h!]
    \centering
    \includegraphics[width=0.5\textwidth]{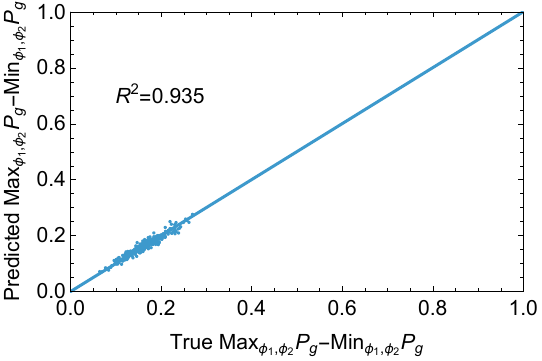}
    \caption{
        {
        \textbf{Quantitative validation of the coherence model.}
        Scatter plot of the predicted maximum interference amplitude versus the true measured amplitude for a wide range of prepared states. The ``True Max" (horizontal axis) is the total contrast ($P_{\text{max}} - P_{\text{min}}$) extracted directly from numerical simulation by scanning the verification phases $\phi_1$ and $\phi_2$. The ``Predicted Max" (vertical axis) is using the magnitudes of the dominant Fourier components of the interference signal, $P_g^{(1)}$, $P_g^{(2)}$, $P_g^{(3)}$, as defined in Eqs.~(D6-D8). The strong linear correlation, with a coefficient of determination $R^2 \approx 0.94$, provides direct quantitative proof that the contrast is largely determined by the underlying multi-pathway quantum coherence.
        }
    }
    \label{fig:validation_plot}
\end{figure}

In the main text, we argue that the interference contrast is a direct signature of the prepared state's coherence. Here, we provide a rigorous quantitative validation of this claim. The total interference signal, $P_g(\phi_1, \phi_2)$, 
%can be decomposed via a Fourier series into components oscillating at different frequencies with respect to the verification phases $\phi_1$ and $\phi_2$. 
{probes the state's \textit{total} coherence, which includes both the qubit-oscillator coherence and the coherence within the oscillator subspaces. 
It is given by 
\begin{align}
   P_g(\phi_1, \phi_2)= P_g^{(1)}+P_g^{(2)}+P_g^{(3)}+c.c.
\end{align}
This signal can be decomposed via a Fourier series into components oscillating at different frequencies}.
As shown in Eqs.~(D6-D8), the amplitudes of these Fourier components, denoted $P_g^{(k)}$, correspond directly to specific coherence terms (i.e., off-diagonal elements) in the state's density matrix. For instance, $P_g^{(1)}$ relates to coherences between Fock states differing by one quantum, $P_g^{(3)}$ relates to coherences between states differing by three quanta, and so on.

The total contrast of the interference fringe, defined as the experimentally measured difference $P_{\text{max}} - P_{\text{min}}$, should therefore be determined by these individual coherence terms. 
%To verify this, we compare the contrast of the randomly sampled probability (``True Max'') with the predicted contrast predicted by the absolute values of the dominant Fourier amplitudes derived from our model (``Predicted Max'').
{To verify this, we checked the correspondence for a wide range of randomly sampled qubit-oscillator parity-entangled states, confirming the generality of this relationship beyond our specific preparation. 
The total contrast (``True Max") was found to be linearly dependent on the magnitudes of the first, second, and third-order coherence terms, $R'^{(1)}=|P_g^{(1)}|$, $R^{(1)}=|P_g^{(1)}|$, $R^{(2)}=|P_g^{(2)}|$ and $R^{(3)}=|P_g^{(3)}|$ (``Predicted Max"), specifically by the approximate formula 
\begin{align}
    \approx 1.73 R'^{(1)}+1.07 R^{(1)}+1.14 R^{(2)}+1.87 R^{(2)}.
\end{align}
}

Figure~\ref{fig:validation_plot} plots this comparison across numerous experimental runs with different preparation sequences. The data points fall tightly along the line $y=x$, demonstrating a near-perfect linear relationship with a coefficient of determination $R^2 \approx 0.94$. 
Averaging over the preparation phases $\bm{\varphi}$ makes the correspondence of the coherence and average contrast very precise.
This result serves as powerful, direct evidence for our central argument: the experimentally observed interference contrast is not merely related to coherence, but is quantitatively determined by the underlying quantum coherence pathways. This validates the use of contrast as a robust and accurate measure of the coherence present in the prepared state.

\section{Longer preparation sequence }
\label{appendix:longer}

\begin{figure}
    \centering
    \includegraphics[width=0.45\textwidth]{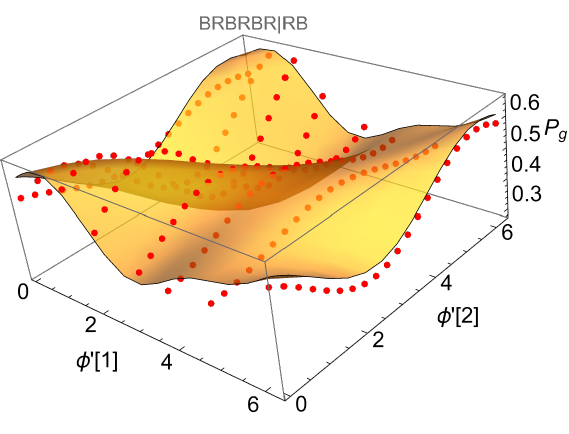}
     \includegraphics[width=0.45\textwidth]{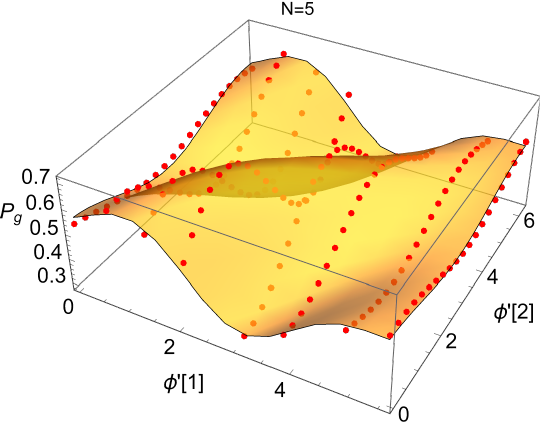}
    \caption{Reconstruction of modulation for $N=6$ (BRBRBR\textbar RB) and $N=10$.}
    \label{fig:PgN3}
\end{figure}

\begin{figure}
    \centering
    \includegraphics[width=1.0\textwidth]{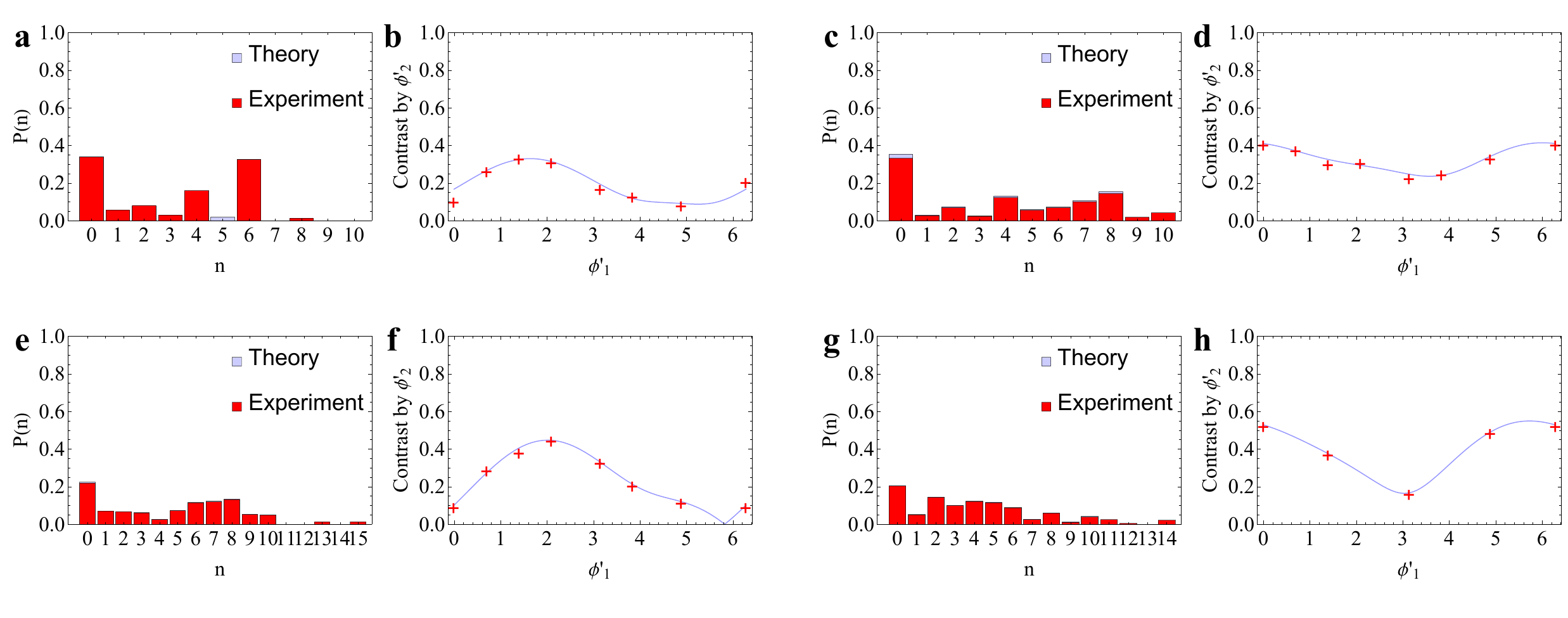}
    \caption{Reconstruction of experimental phonon number distribution P(n) and the contrast by verification pulse phase for $N=6,8,10,12$.}
    \label{fig:PgN6}
\end{figure}

Longer preparation sequences can be reconstructed similarly. In Fig. \ref{fig:PgN3} and \ref{fig:PgN6}, we reconstructed modulation for $N=6$ that shows decoherence factor $w=0.535$ and $N=10$, showing a good agreement.

\section{Contribution to $P_g$ by the off-diagonal elements in the prepared state}
\label{app:offdiag}

{
If the prepared state has the term $\ket{e,n}\bra{g,m}$ in the density matrix, it contributes to $P_g$ after the verification pulses made of a red pulse as:
\begin{align}
    &P_g[\ket{e,n}\bra{g,m}]=\ii \sin[\kappa \sqrt{n+1}]z \cos[\kappa \sqrt{m}]\delta_{n-1,m}.%-\cos[\kappa \sqrt{n+1}]\ii \sin[\kappa \sqrt{m}]z \delta_{n,m-1}.
\end{align}
The only case that gives non-zero contribution is when $n=m+1$. 
Terms with qubit states in $\ket{e}\bra{e}$ or $\ket{g}\bra{g}$ do not have terms with $z$, and thus is not modulating.
If a verification blue pulse acts on the prepared state that has the term $\ket{e,n}\bra{g,m}$ in the density matrix, it contributes to $P_g$ as
\begin{align}
P_g=\ii \sin[\kappa \sqrt{n}]\cos[\kappa \sqrt{m+1}]z^* \delta_{n-1,m}.    
\end{align}
The only non-zero term arises when $n=m+1$.
These results show that only the qubit coherence is contributing to the modulation of $P_g$. 
}

{While single pulses only probe first-order coherences ($|n-m|=1$), the combined RSB-BSB sequence acts as a multi-quantum operator.}
If the prepared state has the term $\ket{e,n}\bra{g,m}$ in the density matrix, it contributes to $P_g$ after the verification pulses made of a red and a blue pulse given as:
\begin{align}
    &P_g[\ket{e,n}\bra{g,m}]=\ii \cos[\kappa \sqrt{n+1}]\sin[\kappa \sqrt{n}]z^*(\cos[\kappa \sqrt{m+1}]\cos[\kappa \sqrt{m}]\delta_{n-1,m}-\sin[\kappa \sqrt{m-1}]z\sin[\kappa \sqrt{m}]z^* \delta_{n-1,m-2}) \nonumber\\
    &+\ii \sin[\kappa \sqrt{n+1}]\cos[\kappa \sqrt{n+2}]z(\cos[\kappa \sqrt{m+1}]\cos[\kappa \sqrt{m}]\delta_{n+1,m}-\sin[\kappa \sqrt{m-1}]z\sin[\kappa \sqrt{m}]z^* \delta_{n+1,m-2})
\end{align}
where $z=e^{\ii \theta}$.
{This allows the sequence to bridge Fock states with higher separations, such as $|n-m|=3$, through intermediate transitions.}
This term is the contribution of qubit coherence and oscillator coherence from terms that differ by odd quanta number, all containing the term $Z=e^{\ii (n-m)\Theta}$. 
{The two-pulse sequence functions as a multi-order filter. By expanding the product of the RSB and BSB unitaries, the resulting POVM contains terms proportional to $|n\rangle\langle n \pm k|$. For our specific pulse areas, the coefficients for $k > 3$ vanish, isolating short-range coherence.} 
Especially, the last term enables the interference by a terms that differ in quanta number by 3. 
From this, we can see that the maximum difference that can be read by such pulses is $n-m=\pm 3$.

If the prepared state has the term $\ket{e,n}\bra{e,m}$, it contributes:
\begin{align}
    &P_g[\ket{e,n}\bra{e,m}]=\ii \cos[\kappa \sqrt{n+1}]\sin[\kappa \sqrt{n}]z^*(-\ii \cos[\kappa \sqrt{m+1}]\sin[\kappa \sqrt{m}]z\delta_{n-1,m-1}-\ii \sin[\kappa \sqrt{m+1}]\cos[\kappa \sqrt{m+1}]z^* \delta_{n-1,m+2}) \nonumber\\
    &+\ii \sin[\kappa \sqrt{n+1}]\cos[\kappa \sqrt{n+2}]z(-\ii\cos[\kappa \sqrt{m+1}]\sin[\kappa \sqrt{m}]z\delta_{n+1,m-1}-\ii\sin[\kappa \sqrt{m+1}]\cos[\kappa \sqrt{m+2}]z^* \delta_{n+1,m+1}).
\end{align}
If the prepared state has the term $\ket{g,n}\bra{g,m}$, it contributes:
\begin{align}
    &P_g[\ket{g,n}\bra{g,m}]=\cos[\kappa \sqrt{n+1}]\cos[\kappa \sqrt{n}]( \cos[\kappa \sqrt{m+1}]\cos[\kappa \sqrt{m}]z^*\delta_{n,m}-\sin[\kappa \sqrt{m-1}]z^*\sin[\kappa \sqrt{m+1}]z^* \delta_{n,m-2}) \nonumber\\
    &+\ii \sin[\kappa \sqrt{n-1}]\sin\kappa \sqrt{n}]z^*( \cos[\kappa \sqrt{m+1}]\cos[\kappa \sqrt{m}]\delta_{n-2,m}-\sin[\kappa \sqrt{m-1}]z^*\sin[\kappa \sqrt{m+1}]z^* \delta_{n-2,m-2}).
\end{align}
The two equations above capture the oscillator coherence, which cannot be measured by a single verification pulse. 
The verification pulse length $\kappa$ can be chosen independently from the preparation pulse length $t$.

\section{Comparison to other methods}

\begin{figure}[tbhp]
    \centering
    \includegraphics[width=1\linewidth]{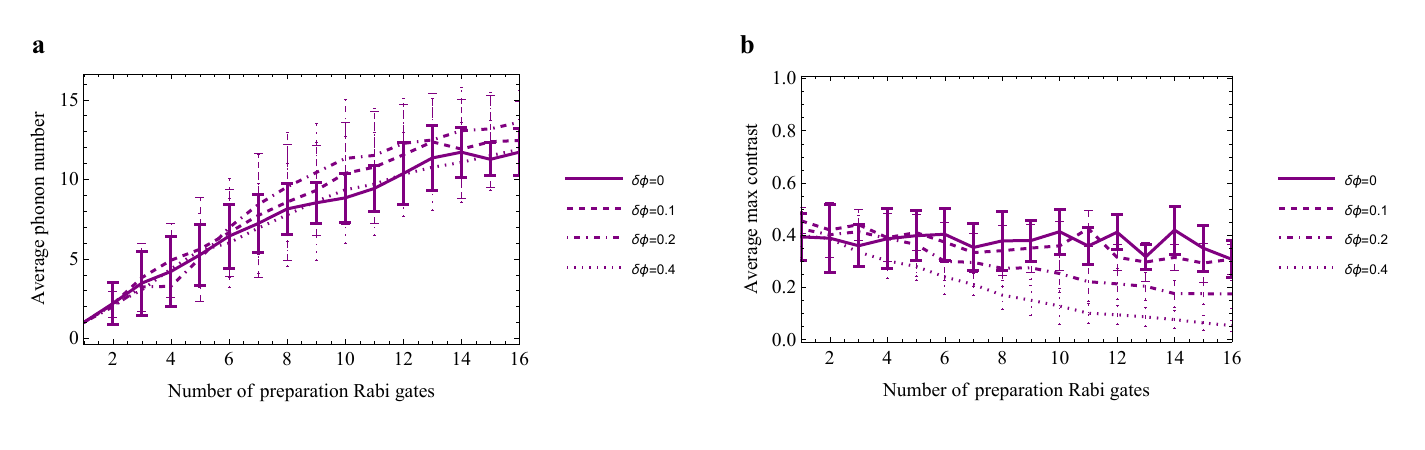}
    \caption{{Rabi gate analysis. \textbf{a} Average phonon number prepared via Rabi gates. It increases similarly as our experiment.
    \textbf{b} Maximum contrast by verification by Rabi gates. The maximum contrast decreases very rapidly by the phase instability in the preparation.}}
    \label{fig:avNmaxcontRabi}
\end{figure}

% Our protocol can be compared to other protocols using different qubit-oscillator interactions.
% For example, Rabi gate of the form $\exp[\ii t (e^{\ii\phi_R}\sigma_+ \hat{a}+e^{-\ii\phi_R}\sigma_- \hat{a}^\dagger+e^{\ii\phi_B}\sigma_+ \hat{a}^\dagger+e^{-\ii\phi_B}\sigma_- \hat{a})]$ can be synthesized digitally by alternating red sideband pulse and blue sideband pulse rapidly, thus providing similarity to our protocol. 
% Both of random phases $\phi_{R,B}$ suffers instability $d\phi$.
% These gates are used instead of the sideband pulses in both preparation and verification.
% In Fig. \ref{fig:avNmaxcontRabi}, such a protocol was investigated.
% By the increased number of preparation Rabi pulses $N$ and phase instability, the max contrast is decreasing. %and it doesn’t seem to have any saturation to a finite value, although a further confirmation is needed. 
% Also the sensitivity toward the phase instability seems slightly larger than our protocol, which was rather robust to upto phase instability $\delta\phi\approx 0.8$. This shows a different behavior from our protocol. The average phonon number is increasing analogously to our protocol. 
Our experiment can be compared to other methods using different qubit-oscillator interactions. For instance, a Rabi gate of the form \(\exp[\ii t (e^{\ii\phi_R}\sigma_+ \hat{a} + e^{-\ii\phi_R}\sigma_- \hat{a}^\dagger + e^{\ii\phi_B}\sigma_+ \hat{a}^\dagger + e^{-\ii\phi_B}\sigma_- \hat{a})]\) can be synthesized digitally by rapidly alternating red and blue sideband pulses. Both gates suffer from phase instability \(d\phi\).
These gates replace sideband pulses in preparation and verification. In Fig. \ref{fig:avNmaxcontRabi}, this analysis with Rabi gates was investigated. With an increasing number of preparation Rabi pulses \(N\) and phase instability, the maximum contrast decreases. Sensitivity to phase instability also appears slightly greater than in our experiment, which is robust to phase instability up to \(\delta\phi \approx 0.8\). This presents a different behavior compared to our experiment. The average phonon number increases analogously in both methods.
When  an arbitrary qubit-oscillator interaction is used for the verification, the contrast  also declines sharply due to phase instability. Therefore, the large contrast under a phase instability is due to the nice property of the sideband verification method.

% \section{Saturation of the average and variance of phonon number}

% The prepared state has increasing average and variance of phonon number very rapidly depending on the number of pulses. 
% However, they soon reaches certain ceiling in the increase. 
% This happens because 

{

\section{Interferometric Fidelity and  Visibility and Contrast}
\label{app:theory_and_sim}

%This appendix provides the theoretical background and simulation results that support the analysis in the main text. 
We first establish the general framework for quantifying interference, then explore the physical conditions that govern fringe visibility, and finally present numerical simulations demonstrating the robustness of high visibility against phase imperfections through practical optimization.

\subsection{General Interferometric Framework}

Interference phenomena in two-path systems, such as the qubit-oscillator states discussed, produce a measurement fidelity $F$ that oscillates with a relative phase $\phi$. 
This dependence can be universally described by the form:
\begin{equation}
F(\phi) = a + b \cos(\phi)
\end{equation}
The parameters $a$ and $b$ are determined by the overlaps of the prepared quantum states and the measurement involved. 
We use contrast and visibility to characterize the quality of the interference fringe:
The contrast is defined as $C = F_{\text{max}} - F_{\text{min}} = 2b$. It measures the absolute probability swing and is directly related to the interference amplitude $b$.
The visibility is given  as $V = (F_{\text{max}} - F_{\text{min}}) / (F_{\text{max}} + F_{\text{min}}) = b/a$. 
%Visibility measures the clarity of the fringe relative to the mean signal $a$. 
A visibility of $V=1$ signifies that the fringe modulates perfectly from zero to its maximum, indicating perfectly balanced interfering pathways, even if the overall contrast is low.
%These metrics are crucial for understanding the coherence of the prepared state. 
%High visibility, in particular, serves as a powerful witness to the balanced nature of the underlying quantum superposition to be shown below.

\subsection{Factors Governing Fringe Visibility}

Our theoretical analysis reveals that fringe visibility is fundamentally tied to the balance between the two interfering quantum pathways. Any asymmetry in the magnitudes of the pathway amplitudes will degrade the visibility below its maximum value of 1.

\subsubsection{The Role of Pathway Balance}
We analyzed several simple conceptual models of general qubit-oscillator Schr\"odinger cat state $\sqrt{w}\ket{e}\ket{\alpha} + e^{\ii\phi}\sqrt{1-w}\ket{g}\ket{-\beta}$:
\begin{itemize}
    \item \textbf{Balanced States:} For states of the form $\frac{1}{\sqrt{2}}(\ket{e}\ket{\alpha} + e^{\ii\phi}\ket{g}\ket{-\alpha})$, the two interfering components have equal magnitude. This leads to $a=b$ and thus a perfect visibility of $V=1$, regardless of the coherent state amplitude $\alpha$ or other phase mismatches.
    For example, if the prepared motional states are the coherent states $\ket{\psi_\mathrm{odd}} = \ket{\alpha}$ and $\ket{\psi_\mathrm{even}} = \ket{-e^{\ii\theta}\alpha}$, where $\theta$ is an unknown phase from the preparation interaction, the visibility remains unity irrespective of $\theta$. 
     In the same example, the unnormalized contrast $C$ would depend explicitly on the phase $\theta$ and could be smaller than unity.
    
    \item \textbf{Amplitude Imbalance:} For a more general state with distinct coherent state components, $\ket{e}\ket{\alpha}$ and $\ket{g}\ket{\beta}$, the visibility is degraded if the amplitudes differ, i.e., $|\alpha| \neq |\beta|$. The interference is imperfect because the contributing state overlaps have unequal magnitudes.
    
    \item \textbf{Weight Imbalance:} If the qubit components themselves are unbalanced, as in $\sqrt{w}\ket{e}\ket{\alpha} + e^{\ii\phi}\sqrt{1-w}\ket{g}\ket{-\alpha}$ with $w \neq 1/2$, the visibility is inherently reduced to $V = 2\sqrt{w(1-w)} < 1$. This is a direct consequence of the unequal probability amplitudes of the two ``arms" of the interferometer.
\end{itemize}

\subsubsection{Restoring Visibility via Optimized Measurement}
Crucially, it is possible to counteract certain types of imbalance to regain the high visibility. 
Our analysis shows that even if an input state is prepared with unbalanced weights ($w \neq 1/2$), perfect visibility ($V=1$) can, in principle, be restored by projecting onto a carefully chosen measurement state. 
Specifically, a measurement state with ``swapped" weights ($w' = 1-w$) perfectly balances the interferometric pathways.

However, this restored visibility comes with a caveat: the absolute contrast remains limited by the initial imbalance. 
The maximum achievable contrast in this optimized scenario is $C_{\text{opt}} = 4w(1-w)$, which is always less than 1 for an unbalanced input. 
This highlights a key distinction: visibility speaks to the \textit{balance} of interference, while contrast speaks to its absolute strength.

\subsection{Numerical Simulation of Robustness to Phase Imperfections}

To connect this theoretical understanding to more realistic experimental conditions, we performed numerical simulations of a preparation-detection sequence where pulse phases are stable but unknown (i.e., random for each sequence realization but fixed within it).

\subsubsection{Optimizing Pulse Areas without Phase Control}
Direct optimization of all pulse phases is experimentally demanding. We therefore investigated a more practical scheme where only the detection pulse \textit{areas} (a proxy for strength and duration) are optimized to maximize interference, while the phases remain uncontrolled. The detection sequence length ($N'$) was set equal to the preparation sequence length ($N$).

\subsubsection{Simulation Results}
The simulations, summarized in Figure \ref{fig:app_sim_results}, reveal a key finding:
\begin{itemize}
    \item \textbf{High visibility can be maintained} across various circuit depths ($N$) by optimizing only the detection pulse areas. This demonstrates that the system is robust against phase imperfections in its ability to produce clear, well-balanced interference fringes.
    \item \textbf{Average contrast remains low} and is highly sensitive to the lack of phase control.
\end{itemize}

\begin{figure}[h!]
    \centering
    % NOTE: Replace 'placeholder.png' with the relevant figure from your simulation report,
    % typically showing <V> and <C> vs. N.
    \includegraphics[width=0.5\textwidth]{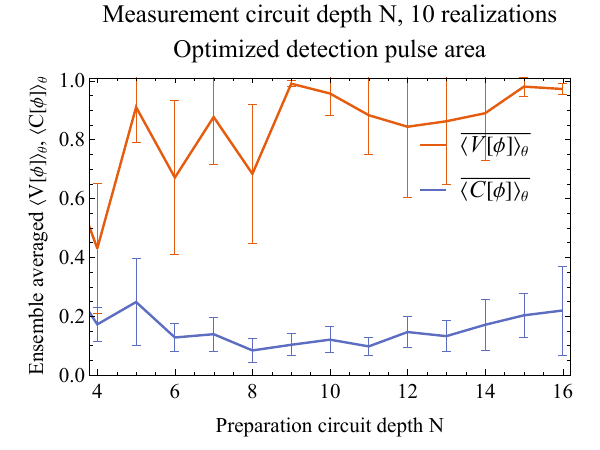}
    \caption{Simulated average visibility $\langle V \rangle$ and contrast $\langle C \rangle$ versus preparation circuit depth $N$. 
    The detection pulses depth were set as equal to $N$.
    For each realization, detection pulse areas were numerically optimized for a sequence with random preparation phases. The results show that high average visibility can be achieved, while average contrast remains low, demonstrating the robustness of visibility to phase imperfections.}
    \label{fig:app_sim_results}
\end{figure}

\subsection{Conclusion}
The combined theoretical and numerical results establish a clear hierarchy: fringe contrast is a fragile metric, sensitive to both state imbalances and phase errors. 
Fringe visibility, on the other hand, is a robust measure of the underlying coherence and optimality of the measurement. 
It is primarily sensitive to the balance of the interfering pathways, and our simulations show that high visibility can be maintained even in the absence of precise phase control by optimizing other accessible experimental parameters such as pulse area. 
%This provides strong support for using high visibility as the primary witness of non-classical interference in our experiment.

}

\section{Experimental setup - Preparation and Measurement of Superposition States}
\label{appendix:experiment}

The experimental verification of the presented scheme is carried out with a single $^{40}\text{Ca}^+$ ion confined in a linear Paul trap with a RF frequency of 29.9~MHz and secular frequencies of $\omega_{x,y,z} \sim 2\pi \times (2,2,1.1)$~MHz in the two radial and axial directions, respectively.
The experimental sequence starts by initialization of the ion to \(|g,0\rangle\) state by the combination of laser cooling and optical pumping pulses. Doppler cooling of all three normal modes is performed using a 397~nm laser red detuned from the $4^2 \rm S_{1/2} \leftrightarrow 4^2 \rm P_{1/2}$ electric dipole transition by $\approx 12$~MHz. The laser is directed at an angle $\alpha \approx 45^{\circ}$ with respect to axial trap direction, within $z-x$~plane. Simultaneously, the 866~nm laser beam, co-propagated along the same direction, is used for reshuffling the population of the \(3^2 \rm D_{3/2}\) manifold back to the cooling transition. A static magnetic field of $\vert \vec{B}\vert = 3.3$~G is applied along the $y$~axis. Both 397~nm and 866~nm laser beams are set to linear polarization oriented perpendicular to the magnetic field.  Sideband cooling is applied on the quadrupole transition $4^2 {\rm S}_{1/2} (m=-1/2) \leftrightarrow 3^2 {\rm D}_{5/2}, m=-5/2$ using a 729~nm laser. A regular sequence of short optical pumping pulses of \(\sigma\)-polarized 397~nm laser and a close-to-resonant 854~nm beam are set to reshuffle the electronic populations back to the $4^2 \rm {\rm S}_{1/2} (m=-1/2)$ level. Both the optical pumping 397~nm beam and the 854~nm repumper propagate along the axial direction. The resulting motional distribution corresponds to \(P_0 \approx 98\%\).

As described in the main part of the article, the generation of large qubit-oscillator superposition states proceeds via N-times sequential iterative excitation of the blue (BSB) and red (RSB) sidebands on the \(\vert 4^2 {\rm S}_{1/2} (m=-1/2) \rangle \leftrightarrow \vert 3^2 {\rm D}_{5/2} (m=-5/2) \rangle\) transition. Each of these laser pulses is set to a constant pulse area given by the temporal length corresponding to \(\eta\Omega_0 t \approx \pi/4\), where  \(\eta\) is the Lamb-Dicke parameter, \(\Omega_0\) is the carrier Rabi frequency. The sequence begins with a resonant BSB, coherently exciting the electronic and motional states into a superposition of \(\vert 0,g \rangle\) and \(\vert 1,e \rangle\). It is followed by successive RSB and BSB pulses, which produce the desired qubit-oscillator state described by Eq.~(3) of the main article.

We investigate the resulting superposition state by three methods, schematically sketched in Fig.~1(c-e). First, we reconstruct the oscillator statistics, second, we analyze qubit-oscillator parity, and third, we perform qubit-oscillator interferometry to analyze coherence of the produced state.

The reconstruction of the oscillator statistics proceeds by transfering the D-state population to the electronic ground state by applying a short pulse of 854~nm laser. Analysis of the time-resolved Rabi oscillations on the BSB transition then allows for motional population estimation by using the model
\begin{equation}
P_g(t) = \frac{1}{2} \left[ 1 + \sum_n P_n \cos\left(\Omega_{n,n\pm 1} t \right) \right]\exp^{-\gamma_n t}
\label{eq_fit}
\end{equation}
where \(P_n\) is the occupation probability of the $n^{\mathrm{th}}$ motional state with estimated \(\Omega_0~= 2\pi ~\times~21.7~\pm~0.03\)~kHz, and \(\gamma_n\) is decay rate which scales with mean phonon number as \(\gamma_n = \gamma_0 (n+1)^{0.7}\) with $\gamma_0$ varying from 0.06 to 0.13~$\textrm{ms}^{-1}$ depending on actual setup conditions. Here, we employed the parametrization of Rabi frequencies beyond the Lamb-Dicke approximation, i.e. \(\Omega_{n,n\pm1} = \Omega_0 \left| \langle n\pm1 \vert e^{i\eta(a + a^\dagger)} \vert n \rangle \right|
\), with \(\eta = 0.0629\). The measurement of the $P_g(t)$ is performed with the electron shelving method, realizing 100 repetitions of the experimental sequence to reduce the impact of the projection noise.

The analysis of qubit-oscillator parity follows a similar sequence, with the main distinction corresponding to the application of the 854~nm pulse prior to Rabi oscillation measurement and an additional post-selection on the electronic state. This enables reconstruction of the statistics of the odd and even oscillator parts independently.  The post-selection is implemented again via electron shelving, that is, state-dependent fluorescence detection on the \(4^2 \rm S_{1/2} \leftrightarrow 4^2 \rm P_{1/2}\) transition. In the post-selection, we systematically retain and analyze only cases where the ion is found in the excited state \(\vert e \rangle\) as in this case, the oscillator statistics is unaffected by photon scattering. To access the ground state population, the post-selection is preceded by an additional carrier \(\pi\)-pulse, flipping the qubit population with minimal impact on the oscillator state. This is followed by measurement of Rabi oscillations on RSB and oscillator statistics reconstruction already described.

Both of these methods give us reliable results of the oscillator statistics, provided we choose sufficient sampling of the Rabi flops. To estimate the statistical uncertainty of the reconstruction we use a Monte-Carlo approach. The input statistical distributions are considered to correspond to the minimal uncertainty of each measured datapoint of $P_g(t)$ given by the quantum projection noise described by the binomial distribution $\sigma_{P_g} = \sqrt{\frac{P_g(1-P_g)}{N}}$. The simulation of uncertainties of resulting parameters using this approach effectively captures the minimal statistical variation of our measurements. A least-squares method is used for the fit of the experimental data using Eq.~\ref{eq_fit}, yielding an initial set of phonon populations. We implement and propagate \(N=100\) random samples of each data point according to its binomial uncertainty.

To analyze the coherence of the prepared qubit-oscillator superposition states, we implement a Ramsey-type interference sequence corresponding to the  single-pulse and two-pulse measurements described in the main text. The latter uses a sequence of an RSB pulse with the controllable phase $\phi_1$ and BSB pulse with the controllable phase $\phi_2$. The first RSB pulse acts on \(\vert g,2n+1 \rangle\leftrightarrow \vert e,2n  \rangle\) transitions, probing qubit-oscillator coherence. The second BSB pulse address \(\vert e,2n \rangle\leftrightarrow \vert g,2n-1  \rangle\), enabling us to investigate the internal oscillator coherence. The method scans both phases $\phi_1, \phi_2$, leading to interference fringes shown in Figs.~\ref{fig:Pevsphi} and \ref{fig:macroscopicitynbarvsN} in the main part of the manuscript. Interference by the single-pulse measurement method works equivalently but uses only the RSB pulse, thus accessing the complex qubit-oscillator coherence. The implementation of those methods in the experiment is as follows: After the preparation pulse sequence, a controllable time delay of \(\tau\) is set to test the decoherence of the prepared superposition states. After that, we apply interference by a single-pulse and two-pulse measurement scheme where we scan $\phi_1$ or $\phi_2$ to obtain an interference fringe pattern. Scanning only one of the phases is sufficient as the coherence test requires only sum or difference of $\phi_1$ and $\phi_2$. This method allows us to probe the qubit-oscillator coherence and internal oscillator coherence even with the presence of an unknown but stable phase shift obtained during the pulse sequence preparation, inversion, and inserted time delay.

Apart from the fixed phase offsets of applied laser pulses, there is a dynamic phase drift caused by residual frequency difference \(\delta \omega\) between the oscillator frequency and the reference oscillator frequency. In addition, trap or laser frequency drifts, and magnetic field fluctuations, correspond to instabilities and decoherence, which are generally unavoidable but slow enough to observe the coherent phenomena on the demonstrated superpositions of up to $\Delta n \approx 10$. Those effects result in a phase evolution of the corresponding superposition state, which scales proportionally to the energy difference of the participating oscillator quanta. Spin coherence time estimated from the Ramsey interferometry by scanning the Ramsey time delay between two carrier \(\pi/2\) pulses corresponds to \(\sim\)8~ms, corresponding to 1/e time. It is primarily limited by laser phase noise and the long-term B-field drifts \(\sim\) 0.2~mG/hr, which, in addition, limit the overall duration of the measurements involving spin motional coherence of the Ramsey contrast but remain stable on the timescales of a few full pulse sequences. Motional coherence, estimated from the motional Ramsey interferometry for an equal superposition of \(\vert 0 \rangle\) and \(\vert 1 \rangle\), defined by the exponential decay of the Ramsey fringe contrast, is \(\sim 100\)~ms. It is predominantly attributed to motional heating, estimated \(\sim\)3 phonons/s and motional dephasing. The coherence decays faster for larger motional superpositions due to their increased sensitivity to drifts of the trapping frequency and faster diffusion of population for higher Fock states due to motional heating. % which results in the corresponding variance of the superposition phase as
%\begin{equation}
%    \Delta \phi^2 \propto n^2 \langle (\delta \nu_{\text{trap}})^2 \rangle \tau^2
%\end{equation}
%The corresponding coherence amplitude measured at a time delay \(\tau\) reduces with the increasing phonon number, as the phase variance scales by \(n^2\), which leads to faster dephasing for motional states involving higher phonon number.

{We note that the presented experimental tests of coherence were conducted over an overall time span of approximately one year, with the complete datasets corresponding to several weeks of continuous data acquisition. Such extended experimental time scales naturally introduce additional systematic variations in laboratory environmental conditions and include small improvements to the experimental apparatus. Consequently, minor discrepancies between different datasets may arise that are not fully captured by the statistically evaluated error bars presented here. 
}

%\end{appendices}

\bibliographystyle{apsrev4-2}
\bibliography{randomcat}

\newpage

\end{document}